\batchmode
\makeatletter
\def\input@path{{"C:/Users/flavi/OneDrive/PhD_PHYSICS/artigos/band projection/"}}
\makeatother
\documentclass[aps,prl,reprint,superscriptaddress,showpacs,longbibliography,floatfix,footnoteinbib]{revtex4-1}
\usepackage[latin9]{inputenc}
\usepackage{color}
\usepackage{float}
\usepackage{amsmath}
\usepackage{amssymb}
\usepackage{graphicx}
\usepackage[pdftex,
 bookmarks=false,
 breaklinks=false,pdfborder={0 0 1},backref=false,colorlinks=true]
 {hyperref}
\hypersetup{pdftitle={Title},
 plainpages=false,pdfpagelabels,linkcolor=blue,urlcolor=blue,citecolor=blue,pdfdisplaydoctitle=true,pdfduplex=DuplexFlipLongEdge}

\makeatletter

\DeclareTextSymbolDefault{\textquotedbl}{T1}
\providecommand{\tabularnewline}{\\}

\usepackage{units}\usepackage{wasysym}

\definecolor{orange}{rgb}{0.50, 0.20, 0.0}

\usepackage{bm}
\usepackage{braket}

\makeatother

\begin{document}
\title{Effective band-projected description of interacting quasiperiodic
systems}
\author{Flavio Riche}
\affiliation{CeFEMA-LaPMET, Physics Departement, Instituto Superior Técnico, Universidade
de Lisboa, Av. Rovisco Pais, 1049-001 Lisboa, Portugal}
\author{Raul Liquito}
\affiliation{Centro de Física das Universidades do Minho e Porto, LaPMET, Departamento
de Física e Astronomia, Faculdade de Ciências, Universidade do Porto,
4169-007 Porto, Portugal}
\author{Bruno Amorim}
\affiliation{Centro de Física das Universidades do Minho e Porto, LaPMET, Universidade
do Minho, Campus of Gualtar, 4710-057, Braga, Portugal}
\affiliation{International Iberian Nanotechnology Laboratory (INL), Av. Mestre
José Veiga, 4715-330 Braga, Portugal}
\author{Eduardo V. Castro}
\affiliation{Centro de Física das Universidades do Minho e Porto, LaPMET, Departamento
de Física e Astronomia, Faculdade de Ciências, Universidade do Porto,
4169-007 Porto, Portugal}
\affiliation{Beijing Computational Science Research Center, Beijing 100084, China}
\author{Pedro Ribeiro}
\affiliation{CeFEMA-LaPMET, Physics Departement, Instituto Superior Técnico, Universidade
de Lisboa, Av. Rovisco Pais, 1049-001 Lisboa, Portugal}
\affiliation{Beijing Computational Science Research Center, Beijing 100084, China}
\author{Miguel Gonçalves}
\affiliation{Princeton Center for Theoretical Science, Princeton University, Princeton,
NJ 08544}
\begin{abstract}
We study the interplay between electronic interactions and quasiperiodicity
in a one-dimensional narrow-band system, focusing on ground-state
and low-energy excitation properties. Using band projection as a low-energy
effective approach, we show that a projection restricted to first
order in the interaction strength fails to reproduce the correlated
phase diagram. This contrasts with the standard success of first-order
band projection in translationally invariant flat-band systems and
highlights the essential role of virtual processes involving remote
bands in quasiperiodic settings. By incorporating second-order interband
contributions perturbatively, we obtain an effective Hamiltonian that
quantitatively reproduces the exact phase diagram previously obtained
using density matrix renormalization group calculations, including
the transition between a Luttinger liquid and a charge-density-wave
phase and the crossover to a quasifractal charge-density-wave regime
at strong quasiperiodicity. We further use this controlled framework
to investigate low-energy neutral excitations and the optical conductivity,
identifying clear dynamical signatures distinguishing the different
phases. Our results establish second-order band projection as a reliable
tool for correlated quasiperiodic narrow-band systems and suggest
a promising route for studying interacting quasiperiodic and moiré
materials beyond one dimension.
\end{abstract}
\maketitle

\section{Introduction}

Quasiperiodic structures can differ fundamentally from periodic crystalline
structures. Incommensurability subtly breaks translational symmetry
in an intrinsic way, leaving long-range correlations between lattice
sites, whose significance has long been recognized in one-dimensional
(1D) systems.

In 1D, quasiperiodicity can dramatically change the electronic properties
relative to the crystalline case, allowing for extended, localized
\cite{aubry1980analyticity,pixley2018weyl,fu2020magic,gonccalves2020incommensurability,yao2019critical,huang2019moire,schreiber2015observation}
or multifractal \cite{gonccalves2023critical,gonccalves2022exact,liu2015localization,liu2021anomalous,wang2021many,wang2020realization}
wave functions. This fundamentally contrasts with the effects of disorder,
that generically localizes the wave function in 1D. The Aubry-André
(AA) model \cite{aubry1980analyticity} is the simplest example of
a 1D quasiperiodic system that describes a quasiperiodicity-induced
transition between an extended and a localized phase. This model can
be experimentally realized by in different platforms, including in
optical and photonic lattices \cite{lahini2009observation,tanese2014fractal,singh2015fibonacci,kohlert2019observation,yao2019critical,liu2021anomalous,An2021}.
Generalizations of the AA can give rise to very rich phase diagrams
with different localization properties, including critical phases
with multifractal eigenstates \cite{liu2015localization,liu2021anomalous,gonccalves2024entanglement,ganeshan2015nearest},
and mobility edges between extended, localized and critical phases
\cite{ribeiro2013strongly,wang2020one,biddle2011localization,deng2019one,gonccalves2023critical,liu2021anomalous,gonccalves2021hidden,luschen2018single}.

The interplay between quasiperiodicity and interactions is a challenging
problem that has had recent important developments in 1D \cite{vu2021moire,gonccalves2024incommensurability,gonccalves2024short,riche2024fractal,yao2024mott,zhang2025quasi,zhang2025reentrant}.
However, the relevance of this interplay in higher dimensions, and
in particular in moiré systems - where quasiperiodicity can also induce
non-trivial localization properties at the single-particle level \cite{fu2020magic,gonccalves2020incommensurability,chou2020magic}
- remains an open question.

In our previous research \cite{gonccalves2024incommensurability}
we showed that quasiperiodicity can be responsible for the emergence
of new correlated phases in a 1D narrow-band system, using DMRG to
obtain the full exact phase diagram. Extending our analysis to higher-dimensional
systems is however significantly more challenging because DMRG and
other exact methods such as exact diagonalization are restricted to
small system sizes. Additionally, calculating dynamic properties,
such as the optical conductivity, becomes particularly challenging
with DMRG, even in one dimension, due to the need of capturing excited
states. With this motivation, we explore the band projection (BP)
technique as a possible route to study interacting quasiperiodic systems,
using the model in Ref.$\,$\cite{gonccalves2024incommensurability}
as a test bed.

Projection onto flat or narrow bands is an approach that dates back
to the study of the fractional quantum Hall effect \cite{girvin2019modern}.
More recently, it has become a standard approach to study the low-energy
physics of interacting moiré systems \cite{andrei2020graphene,li2025quantum}.
The band projection technique has also been applied to investigate
fractional Chern insulators, where the projection of interactions
on a topological flatband provides an effective description of emergent
fractionalized phases \cite{wang2025fractional,liu2012fractional,neupert2011fractional,bergholtz2013topological}.
While it is possible to successfully capture experimentally observed
correlated phases by projecting the Hamiltonian into a single narrow-band,
it is only well justified if the intra-band gap is much larger than
all other energy scales. Away from this extreme regime, remote bands
can also play an important role \cite{sodemann2013landau,herviou2023possible,jia2024moire,yu2024fractional,yu2025moire,abouelkomsan2024band,xu2024maximally,li2025multiband,shen2024stabilizing}.
Moreover, even if band projection methods have been extensively explored
in translationally invariant systems, a comprehensive assessment of
their validity and effectiveness when translational symmetry is explicitly
broken is currently lacking. Answering these questions is of crucial
importance for exploring the physics of moiré systems, where BP could
provide a controlled route to incorporating the interplay between
quasiperiodicity and interactions exactly.

In this work, we apply the band projection technique in a 1D narrow-band
quasiperiodic system, using the model in Ref.$\,$\cite{gonccalves2024incommensurability}
as a test bed for quasiperiodic systems. We find that naive projection
onto the narrow band fails to reproduce any of the ground-state phase
transitions previously obtained with DMRG in a numerically exact manner,
whereas perturbatively including remote-band effects up to second
order in the interaction strength captures both the transitions between
the exact Luttinger liquid (LL) and charge-density-wave (CDW) phases,
as well as the exotic quasi-fractal regime found in Ref.$\,$\cite{gonccalves2024incommensurability}.
Building on the success of this approach to describe the ground-state
phase diagram, we extend our analysis beyond ground-state physics,
investigating dynamical properties of the model, with special focus
on the optical conductivity.

The paper is structured as follows: Section II outlines the BP method
and the physical quantities that we will be studying throughout the
manuscript; Section III presents the results for the ground-state
phase diagram and low-energy neutral excitations; and section IV summarizes
our main findings. Additional results, details on the BP technique
and the computation of the optical conductivity are provided in the
Appendices.

\section{Model and methods}

We consider a 1D lattice model of spinless fermions with nearest-neighbor
hoppings modulated quasiperiodically\cite{liu2015localization,gonccalves2024incommensurability}
and repulsive nearest-neighbor interactions ($U>0$), given by:

\begin{equation}
H=H_{0}+V=-\sum_{r=0}^{N-1}t_{r}\left(c_{r}^{\dagger}c_{r+1}+H.c.\right)+U\sum_{r=0}^{N-1}n_{r}n_{r+1},\label{eq:1}
\end{equation}
where $H_{0}$ and $V$ are respectively the single-particle and interacting
parts of the Hamiltonian, $c_{r}^{\dagger}$ is the creation operator
at site $r$, $t_{r}=1+V_{2}\cos[2\pi\tau(r+1/2)+\phi]$ is the nearest-neighbor
hopping, with modulation strength $V_{2}$ and period $\tau^{-1}$,
$\phi$ is the phase shift and $U$ is the interaction strength. We
implement twisted boundary conditions through the substitution $t\rightarrow te^{i\theta/N},$where
$\theta$ is the phase twist.

\begin{figure}[h]
\centering{}\includegraphics[scale=0.42]{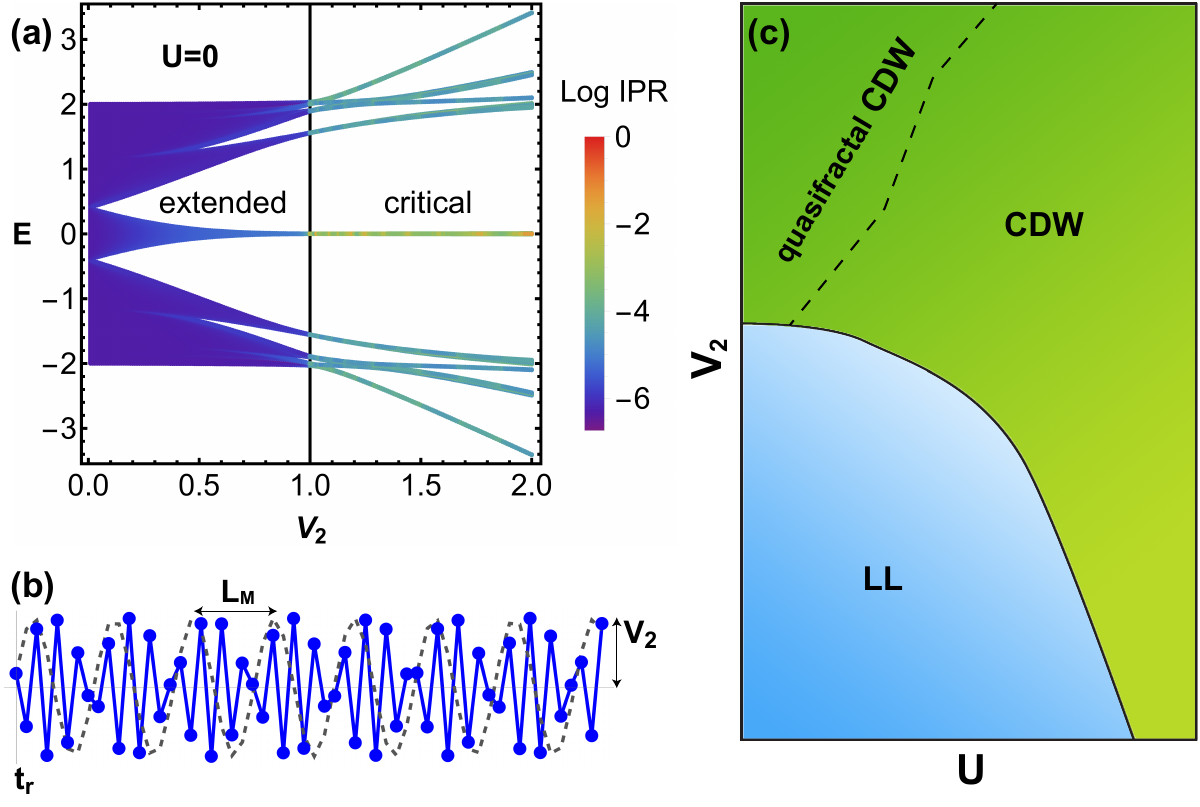}\caption{(a) Single-particle IPR of the model given by Eq. \ref{eq:1} in the
non-interacting limit as a function of the quasiperiodic potential
strength, $V_{2},$ and energy, $E$, for $L=842$, $\tau=475/842$
and $\phi=0$. (b) Hopping modulation of the model, which creates
a moiré pattern of length $L_{M}\simeq7.8$. (c) Phase diagram of
the model in the plane of interaction strength $U$ and intensity
of quasiperiodic hoppings $V_{2}$, adapted from Ref.$\,$\cite{gonccalves2024incommensurability}.
\protect\label{fig:1}}
\end{figure}

When $\tau$ is an irrational number, translational symmetry is broken,
and the system is quasiperiodic. In order to reduce the finite size
effects, we work with rational approximants, $\tau\simeq\tau_{N',N}=N'/N$,
where $N$ is the system size, corresponding to a single unit cell.
In our work we choose $\tau=\sqrt{\pi}^{-1}\simeq0.5642$, which allows
the formation of a nearly flatband (FB) close to $V_{2}=1$, as shown
in Fig. \ref{fig:1}(a), associated with a moiré pattern of length
$L_{M}=(\tau-1/2)^{-1}/2\simeq7.8$, shown in Fig. \ref{fig:1}(b).
It is important to note that because of quasiperiodicity, there is
no longer any notion of Bloch bands. However, it is possible to identify
in Fig. \ref{fig:1}(a) a cluster of states around $E=0$, that we
will refer to as a narrow-band or nearly flatband. As $\tau$ gets
closer to $1/2$, the band becomes considerably flatter, with an increasing
$L_{M}$\cite{liu2015localization,gonccalves2024incommensurability},
in which case the effect of interactions is expected to be enhanced.

The non-interacting model ($U=0$) hosts extended states for $V_{2}<1$
and critical states for $V_{2}\geq1$. Once interactions are turned
on, the model displays Luttinger liquid (LL) and charge density wave
(CDW) phases. For $V_{2}\geq1$ there is a crossover between the CDW
and a quasifractal-charge density wave (qf-CDW), that develops at
smaller $U$. In fact, the quasiperiodic-induced critical phase at
$V_{2}\geq1$ is unstable towards the qf-CDW regime at any finite
$U$. In contrast, when the system is periodic (i.e. $\tau$ is chosen
to be a rational), the small $U$ phase is a LL. While in a truly
fractal-CDW phase the number of ordered wavevectors contributing to
the charge modulations would diverge, in the qf-CDW regime the numerical
results indicate that the number of wavevectors is finite, although
strongly increasing in this regime \cite{gonccalves2024incommensurability}.
The phase diagram in the parameter space of $V_{2}$ and $U$ is schematically
shown in Fig. \ref{fig:1}(c).

Solving Eq. \ref{eq:1} using full exact diagonalization (ED) is unpractical
due to the exponential growth of the Hilbert space, posing significant
limitations to accessing the thermodynamic limit. To overcome this
challenge, we employ the BP technique.

We work at half-filling in an interaction regime that is strong compared
to the width of the narrow band ($W$) and weak compared to the bandgap
($\Delta$), $W\ll U\ll\Delta$. For $V_{2}=0.9$, we have $W\simeq5\times10^{-3}$
and $\Delta\simeq1.5$ . Throughout the manuscript, we focus on $U\leq0.6$.
In this case, the interaction-induced mixing between the FB and the
remote bands can be treated perturbatively. In the absence of interband
excitations, the states in the lower band can be considered as \textquotedbl frozen\textquotedbl ,
since the band is fully occupied and intraband excitations are Pauli-blocked,
while the upper band remains energetically inaccessible and thus completely
empty. For this reason, we label such bands as occupied (OC) and empty
(E).

We start by defining the band creation and annihilation operators

\begin{equation}
d_{i}^{\dagger}=\sum_{r}\langle r|i\rangle c_{r}^{\dagger},\ \:\;d_{i}=\sum_{r}\langle i|r\rangle c_{r},\label{eq:2}
\end{equation}
where $d_{i}^{\dagger}$ creates a particle in level $i$. In the
following, we label $d_{i}^{\dagger}=o_{i}^{\dagger}$, $d_{i}^{\dagger}=p_{i}^{\dagger}$,
and $d_{i}^{\dagger}=e_{i}^{\dagger}$ for $i\in\text{OC, FB, E}$
respectively. Plugging Eq. \ref{eq:2} in Eq. \ref{eq:1} leads to
the first-order (linear in $U$) projected Hamiltonian:
\begin{align}
H_{BP}^{(1)} & =H_{0}+H_{\text{FB-OC}}+H_{\text{FB-FB}}=\sum_{i\in hf}\varepsilon_{i}p_{i}^{\dagger}p_{i}+\nonumber \\
 & 4\sum_{\substack{i,n\ \in\ \text{FB}\\
j\ \in\ \text{OC}
}
}V_{ijjn}^{A}p_{i}^{\dagger}p_{n}+\sum_{\substack{i,j\ \in\ \text{FB}\\
m,n\ \in\ \text{FB}
}
}V_{ijkl}^{A}p_{i}^{\dagger}p_{j}^{\dagger}p_{m}p_{n},\label{eq:4}
\end{align}
where $\varepsilon_{i}$ are the single-particle eigenenergies obtained
from $H_{0}$, $V_{ijmn}^{A}=U/4\left[V_{ijmn}-V_{jimn}-V_{ijnm}+V_{jinm}\right]$
due to antisymmetrization. $H_{\text{FB-OC}}$ is the one-body interacting
term arising due to ``potential scattering'' from electrons in the
occupied band and $H_{\text{FB-FB}}$ is the two-body interacting
term projected into the FB space. Second-order corrections to Eq.$\,$\ref{eq:4}
are obtained by accounting for virtual interband excitations. The
full second-order Hamiltonian can be written as

\begin{equation}
H_{BP}^{(2)}=P_{0}H_{0}P_{0}+P_{0}VP_{0}+\frac{P_{0}V\overline{P}VP_{0}}{E-H_{0}},\label{eq:5}
\end{equation}
where $P_{0}$ is the projection operator onto the unperturbed subspace
that is spanned by states with no holes in the occupied band and no
electrons in the empty band, and $\overline{P}=\mathbb{I}-P_{0}$.

\begin{figure}[h]
\centering{}\includegraphics[scale=0.5]{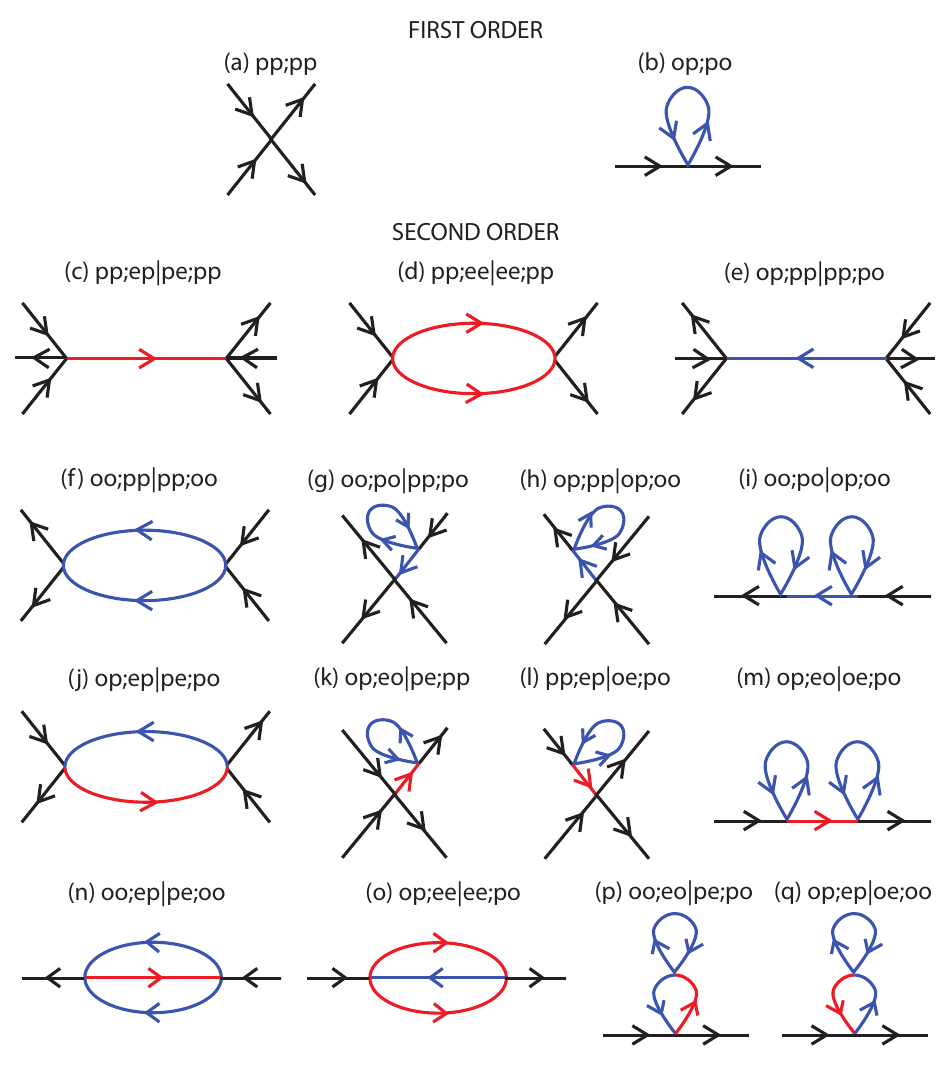}\caption{Feynman diagrams for the effective Hamiltonian. FB, OC and E lines
are depicted in black, blue and red, respectively. Time flows horizontally
to the right. \protect\label{fig:2}}
\end{figure}

Fig. \ref{fig:2} shows the Feynman diagrams for the terms arising
from second-order band projection, where black, blue and red lines
correspond FB, OC and E states, respectively. For simplicity, we use
the notation $ab;cd=V_{\bar{i}\bar{j}kl}^{A}a_{\bar{i}}^{\dagger}b_{\bar{j}}^{\dagger}c_{k}d_{l}$,
with implicit Einstein's notation.

We now re-write the full second-order effective Hamiltonian by combining
all terms after normal ordering:
\begin{align}
H_{BP}^{(2)} & =\varepsilon_{i}p_{i}^{\dagger}p_{i}+V_{in}^{1b}p_{i}^{\dagger}p_{n}+V_{ijmn}^{2b}p_{i}^{\dagger}p_{j}^{\dagger}p_{m}p_{n}\nonumber \\
 & +V_{ijklmn}^{3b}p_{i}^{\dagger}p_{j}^{\dagger}p_{k}^{\dagger}p_{l}p_{m}p_{n},\label{eq:6}
\end{align}
where $V_{in}^{1b}$, $V_{ijmn}^{2b}$ and $V_{ijklmn}^{3b}$ are
the one-body, two-body and three-body interacting terms in the FB
space. A detailed derivation of Eq. \ref{eq:6} is given in Appendix
A.

We now describe the different quantities that we used to characterize
the ground-state phase diagram. To diagnose phase transitions and
crossovers, we use the fidelity susceptibility, defined as \cite{gu2010fidelity}:

\begin{equation}
\chi_{F}=\text{\ensuremath{\lim\limits_{\delta U \to0}}}\left(-\frac{2\text{ln}F}{N\delta U^{2}}\right),\label{eq:7}
\end{equation}
 where $F=|\langle\psi(U)|\psi(U+\delta U)\rangle|$ is the fidelity
and $\psi(U)$ is the many-body ground state for a given interaction
strength $U$.

In order to distinguish between gapless and gapped phases, we study
the charge gap:

\begin{equation}
\Delta_{C}=E(N_{p}+1)+E(N_{p}-1)-2E(N_{p}),\label{eq:8}
\end{equation}
where $E(N_{p})$ is the ground state energy for a system with $N_{p}$
particles. At half-filling, $N_{p}=N/2$.

To explore the formation of charge order, we define the real-space
density fluctuations as $\langle\delta n_{r}\rangle=\langle n_{r}\rangle-1/2$
and their corresponding Fourier transform $\langle\delta n_{k}\rangle=1/\sqrt{N}\sum_{r=0}^{N-1}e^{-i2\pi kr/N}\langle\delta n_{r}\rangle$.
Particle-hole symmetric \cite{gonccalves2024incommensurability} at
half-filling mposes $\langle\delta n_{r}\rangle=0$. Thus, in order
to used $\langle\delta n_{k}\rangle$and a diagnostic tool, we add
a symmetry breaking field to $H_{0}$ in Eq. \ref{eq:1} $\varepsilon c_{0}^{\dagger}c_{0}$,
choosing $\varepsilon=2$. In the gapless phase, the boundary fluctuations
induced by the edge field are non-extensive and we have $\langle\delta n_{k}\rangle\propto N^{-1/2}$
for any $k$. On the other hand, in the CDW phase $\varepsilon$ selects
a symmetry-broken CDW state with density fluctuations that scale as
$\langle\delta n_{k}\rangle\propto N^{1/2}$, with all the wavevectors
$k$ connected to that state. A simple way to quantitatively distinguish
these phases is through the inverse participation ratio (IPR) of the
average of the density fluctuations in momentum space, defined as:
\begin{equation}
\text{IPR}_{K}(\langle\delta\boldsymbol{n}\rangle)=\frac{\sum_{k=0}^{N-1}|\langle\delta n_{k}\rangle|^{4}}{\left(\sum_{k=0}^{N-1}|\langle\delta n_{k}\rangle|^{2}\right)^{2}}.\label{eq:9}
\end{equation}
In the thermodynamic limit, $N\rightarrow\infty$, $\text{IPR}_{K}\sim N^{-1}$for
the LL phase, while $\text{IPR}_{K}\sim N^{0}$ for the CDW phase
\cite{gonccalves2024incommensurability}.

Beyond the ground-state, we also explore low-lying excitations through
the many-body density of states (DoS):
\begin{align*}
\ensuremath{\rho(E)} & =\frac{1}{N}\sum_{\nu=1}^{N}\delta_{\sigma}(E-E_{\nu})
\end{align*}
where $\nu$ labels excited many-body states and $\delta_{\sigma}\simeq\exp\left(-\omega^{2}/2\sigma^{2}\right)/\sqrt{2\pi}\sigma$
is the Gaussian broadening of the delta-function with width $\sigma$,
that we take to be of the order of the mean-level spacing.

With the aim of exploring the system\textquoteright s dynamical properties,
we compute the regular component of the optical conductivity, within
the framework of linear-response theory \cite{shastry1990twisted,stafford1991finite,stafford1993scaling,fye1991drude,millis1990interaction,moreo1990optical,shao2021using}:

{\small
\begin{align}
\sigma_{reg}(\omega) & =\frac{i}{\omega+i\eta}\left[\frac{2}{N}\sum_{\nu\neq0}\frac{|\langle\nu|J|0\rangle|^{2}}{\Delta E_{0}^{\nu}}\right.\nonumber \\
 & \left.+\frac{1}{N}\sum_{\nu\neq0}\frac{|\langle\nu|J|0\rangle|^{2}}{\omega-\Delta E_{0}^{\nu}+i\eta}-\frac{1}{N}\sum_{\nu\neq0}\frac{|\langle\nu|J|0\rangle|^{2}}{\omega+\Delta E_{0}^{\nu}+i\eta}\right],\label{eq:11}
\end{align}
}where $J=-i\sum_{r=0}^{N-1}t_{r}(c_{r}^{\dagger}c_{r+1}-c_{r+1}^{\dagger}c_{r})$
is the current operator, $\Delta E_{0}^{\nu}=E_{\nu}-E_{0}$ is the
energy shift between the ground state and the excited state and $\omega$
is the frequency of the vector field, $A(\omega)$, driving the system's
response. This probing frequency is assumed to be smaller that the
inter-band gap, in which case The intraband transitions can be computed
within the BP formalism, as shown in Appendix , where the the current
term, $\sum_{\nu\neq0}|\langle\nu|J|0\rangle|^{2}$, yields:

\begin{widetext}

\begin{align}
\sum_{\nu\neq0}|\langle\nu|J|0\rangle|^{2} & =\sum_{\nu\neq0}\langle0|\sum_{j,k\in\text{FB}}-i\sum_{r=0}^{N-1}t_{r}\left[\langle j|r\rangle\langle r+1|k\rangle-\langle j|r+1\rangle\langle r|k\rangle\right]p_{j}^{\dagger}p_{k}|\nu\rangle\times\nonumber \\
 & \langle\nu|i\sum_{l,m\in\text{FB}}t_{r}\left[\langle l|r+1\rangle\langle r|m\rangle-\langle l|r\rangle\langle r+1|m\rangle\right]p_{l}^{\dagger}p_{m}|0\rangle.\label{eq:12}
\end{align}

\end{widetext}

Further details on the derivation of Eq. \ref{eq:11}, computation
of its real part and BP implementation are presented in Appendix C.

The properties of the system's dynamical response are also explored
using the spectral weight of the regular part of the optical conductivity
in the $\omega\ll\Delta$ regime:

\begin{equation}
K(\Omega)=\text{\ensuremath{\sum_{\omega<\Omega}\frac{2}{\pi}}}\mathfrak{Re}\left[\sigma_{reg}(\omega)\right]\delta\omega,\label{eq:13}
\end{equation}
where $\Omega$ covers the energy spectrum of the excited states in
the FB.

Finally, for reference, we provide the system sizes and rational approximants
used for all the calculations in Table \ref{tab:1}.

\begin{table}
\begin{tabular}{|c|c|c|c|c|c|c|}
\hline 
$\boldsymbol{N}$ &
$30$ &
$44$ &
$62$ &
$76$ &
$94$ &
$108$\tabularnewline
\hline 
\hline 
$\boldsymbol{\tau}$ &
$17/30$ &
$25/44$ &
$35/62$ &
$43/76$ &
$53/94$ &
$61/108$\tabularnewline
\hline 
\end{tabular}

\caption{Sizes and rational approximants used for numerical computations. \protect\label{tab:1}}
\end{table}

\section{Results and discussion}

We begin by benchmarking the BP approach against the numerically exact
results previously obtained using DMRG in Ref.$\,$\cite{gonccalves2024incommensurability}.
Our first key finding is that a projection restricted to first order
in the interaction strength fails to reproduce any of the phase transitions
of the full model. This result is non-trivial and, at first sight,
surprising: first-order band projection is routinely employed in the
context of fractional quantum Hall systems and other interacting models
with well-separated bands, where it typically captures the correct
qualitative structure of the correlated phases \cite{parameswaran2013fractional,huang2024self,lin2025fractional,qi2011generic,bergholtz2013topological}.

A closer analysis reveals that, in the present quasiperiodic narrow-band
system, this expectation breaks down. As shown explicitly in Appendix
D, a specific second-order virtual process involving remote bands
provides a qualitatively essential contribution, without which the
effective Hamiltonian fails to capture the interaction-driven instabilities
observed in the full DMRG treatment.

In the following, all results are therefore obtained using the second-order
effective Hamiltonian, Eq. (5), which incorporates these virtual interband
excitations in a controlled manner.

\subsection{Ground-state phase diagram}

\begin{figure}[h]
\centering{}\includegraphics[scale=0.26]{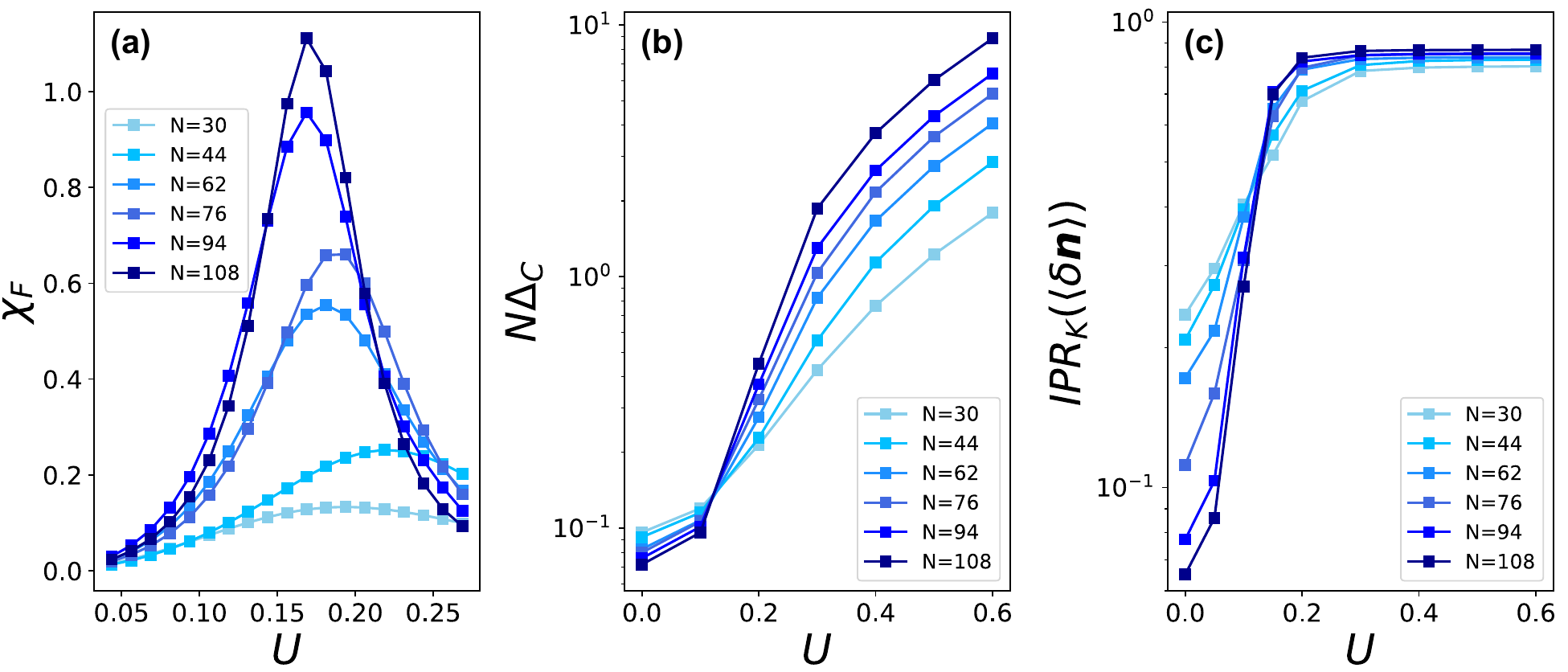}\caption{LL-CDW phase transition determined by the finite-size scaling analysis
of: (a) the fidelity susceptibility ($\chi_{F}$), defined by Eq.
\ref{eq:7} with $\delta U=1.25\times10^{-2}$; (b) the charge gap
($\Delta_{C}$), defined by Eq. \ref{eq:8}; (c) the IPR of the average
of the density fluctuations in momentum space ($IPR_{K}(\langle\delta\boldsymbol{n}\rangle)$),
defined by Eq. \ref{eq:9}. Results are averaged over 10 random configurations
of $\phi$ and $\theta$.\protect\label{fig:3}}
\end{figure}

We first consider the case where interactions are introduced in the
extended single-particle regime, focusing on $V_{2}=0.9$. Figure$\,$\ref{fig:3}(a)
shows the fidelity susceptibility $\chi_{F}$ as a function of interaction
strength $U$ for several system sizes. A pronounced peak develops
at a critical interaction $U=U_{c}$, with a height that grows with
system size, signaling a quantum phase transition.

The nature of the phases on either side of this transition is clarified
by examining the charge gap $\Delta_{C}$ (Fig.$\,$\ref{fig:3}(b{]}),
and the momentum-space inverse participation ratio of the density
fluctuations, $\mathrm{IPR}_{K}(\langle\delta n\rangle)$ {[}Fig.
\ref{fig:3}(c){]}. For $U<U_{c}$, both quantities decrease with
increasing system size, consistent with a gapless and uncorrelated
phase, which we identify as a LL. In contrast, for $U>U_{c}$ both
$\Delta_{C}$ and $\mathrm{IPR}_{K}(\langle\delta n\rangle)$ approach
finite values in the thermodynamic limit, indicating the onset of
a gapped CDW phase with long-range order.

At finite quasiperiodic modulation, the LL--CDW transition is expected
to be of the Berezinskii--Kosterlitz--Thouless (BKT) type, as in
the periodic case. However, the structure of the ordered phase differs
markedly from that at $V_{2}=0$. In the absence of quasiperiodicity,
the CDW is characterized by a single ordering wavevector $k=\pi$
\cite{cazalilla2011one}, whereas for $V_{2}\lesssim1$ the charge
order contains contributions from multiple wavevectors inherited from
the quasiperiodic modulation, in agreement with earlier DMRG results
$\,$\cite{gonccalves2024incommensurability}.

\begin{figure}[h]
\centering{}\includegraphics[scale=0.45]{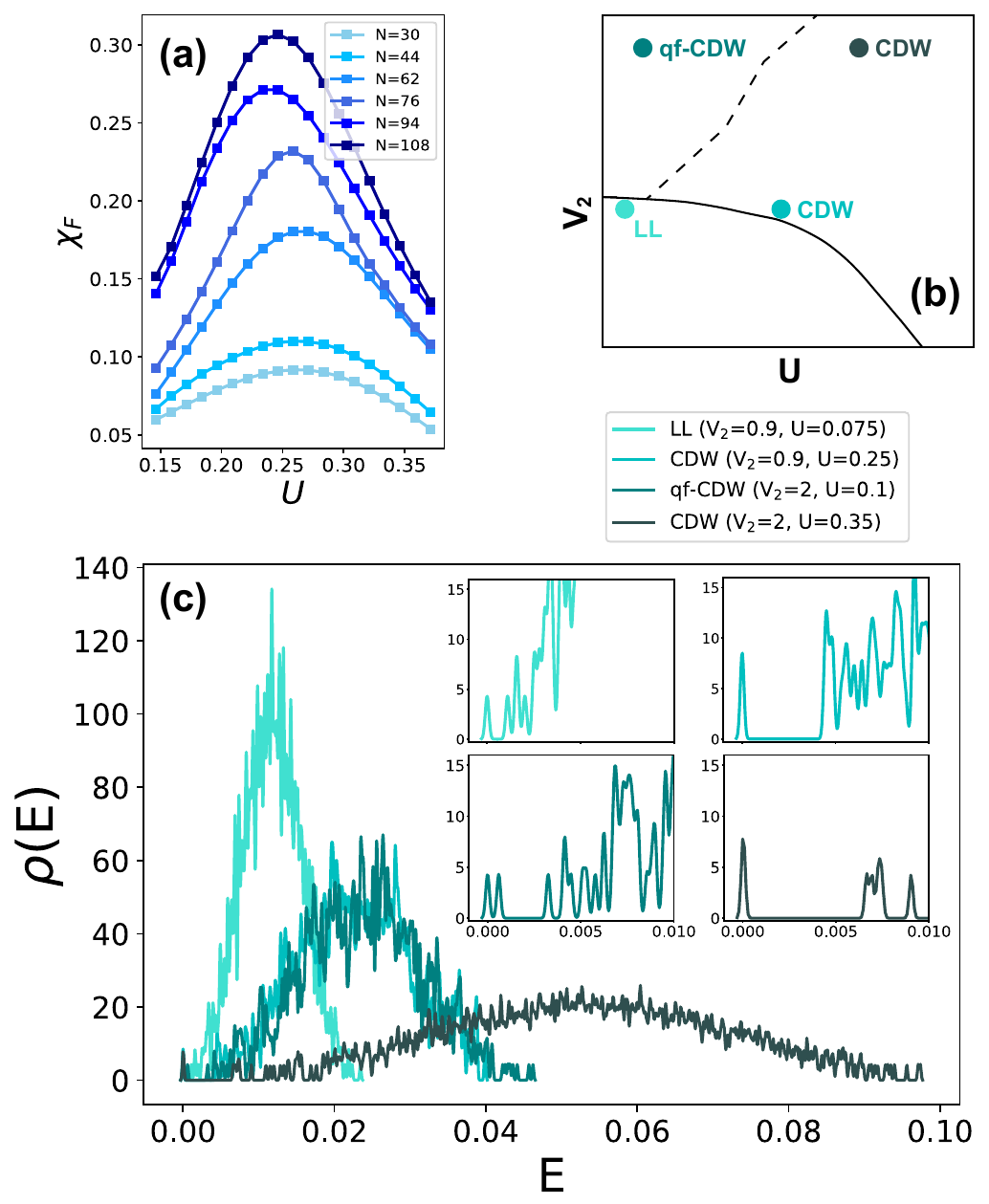}\caption{(a) Fidelity susceptibility ($\chi_{F}$) for $V_{2}=2$. (b) Schematic
phase diagram indicating the phase space points that are used for
computing the density of states and the optical conductivity in panel
(c). (c) Density of states computed with a Gaussian width of $\sigma=10^{-4}$
and $N=94$ for the following phase-space points: $V_{2}=0.9$, $U=0.075$
(LL); $V_{2}=0.9$, $U=0.25$ (CDW); $V_{2}=2$, $U=0.1$ (qf-CDW);
$V_{2}=2$, $U=0.275$ (CDW). Inset shows a close-up at low energies.}
\label{fig:4}
\end{figure}

We now turn to the strongly quasiperiodic regime, $V_{2}\ge2$, where
the non-interacting system hosts critical single-particle states.
Figure$\,$\ref{fig:4}(a) shows the fidelity susceptibility as a
function of $U$. In contrast to the sharp LL--CDW transition observed
at $V_{2}=0.9$, here $\chi_{F}$ displays a broad peak around $U\simeq0.25$.
As shown in Ref.$\,$\cite{gonccalves2024incommensurability}, this
feature corresponds to a crossover between a quasifractal charge-density-wave
(qf-CDW) regime at weak interactions and a more conventional CDW phase
at larger $U$. Although finite-size effects prevent a clear saturation
of $\chi_{F}$ in the present system sizes, the markedly smoother
behavior of the peak already distinguishes this crossover from a true
phase transition.

Together, these results demonstrate that second-order band projection
accurately reproduces the full ground-state phase diagram of the model,
including both the LL--CDW transition in the extended regime and
the qf-CDW--CDW crossover in the critical regime.

\subsection{Low-energy excitation spectrum}

Having established the validity of the BP approach at the level of
ground-state properties, we now turn to low-energy excitations, which
were not accessible in the original DMRG study. Excited states are
obtained by exact diagonalization within the projected narrow-band
Hilbert space.

Figure \ref{fig:4}(c) shows the many-body density of states (DoS)
for representative points in the phase diagram, indicated schematically
in Fig. \ref{fig:4}(b). In the LL phase ($V_{2}=0.9$, $U=0.075$),
the DoS extends continuously to zero energy, reflecting the presence
of gapless neutral excitations. Upon entering the CDW phase ($V_{2}=0.9$,
$U=0.25$), a clear excitation gap opens between the two-fold degenerate
ground-states and the excited state band, consistent with the breaking
of translational symmetry and the formation of long-range charge order.

In the strongly quasiperiodic regime, the qf-CDW phase ($V_{2}=2$,
$U=0.1$) exhibits qualitatively different behavior. While the system
remains ordered, the neutral gap becomes strongly suppressed and the
twofold ground-state degeneracy associated with CDW order is weakly
lifted. This is consistent with the picture of a quasifractal ordered
state, where charge modulations involve a large --- though finite
--- number of incommensurate wavevectors and finite-size effects
are particularly pronounced. At larger interaction strength ($V_{2}=2$,
$U=0.275$), the system crosses over into a conventional CDW phase,
and a robust neutral gap is recovered.

\subsection{Optical conductivity}

\begin{figure}[h]
\centering{}\includegraphics[scale=0.52]{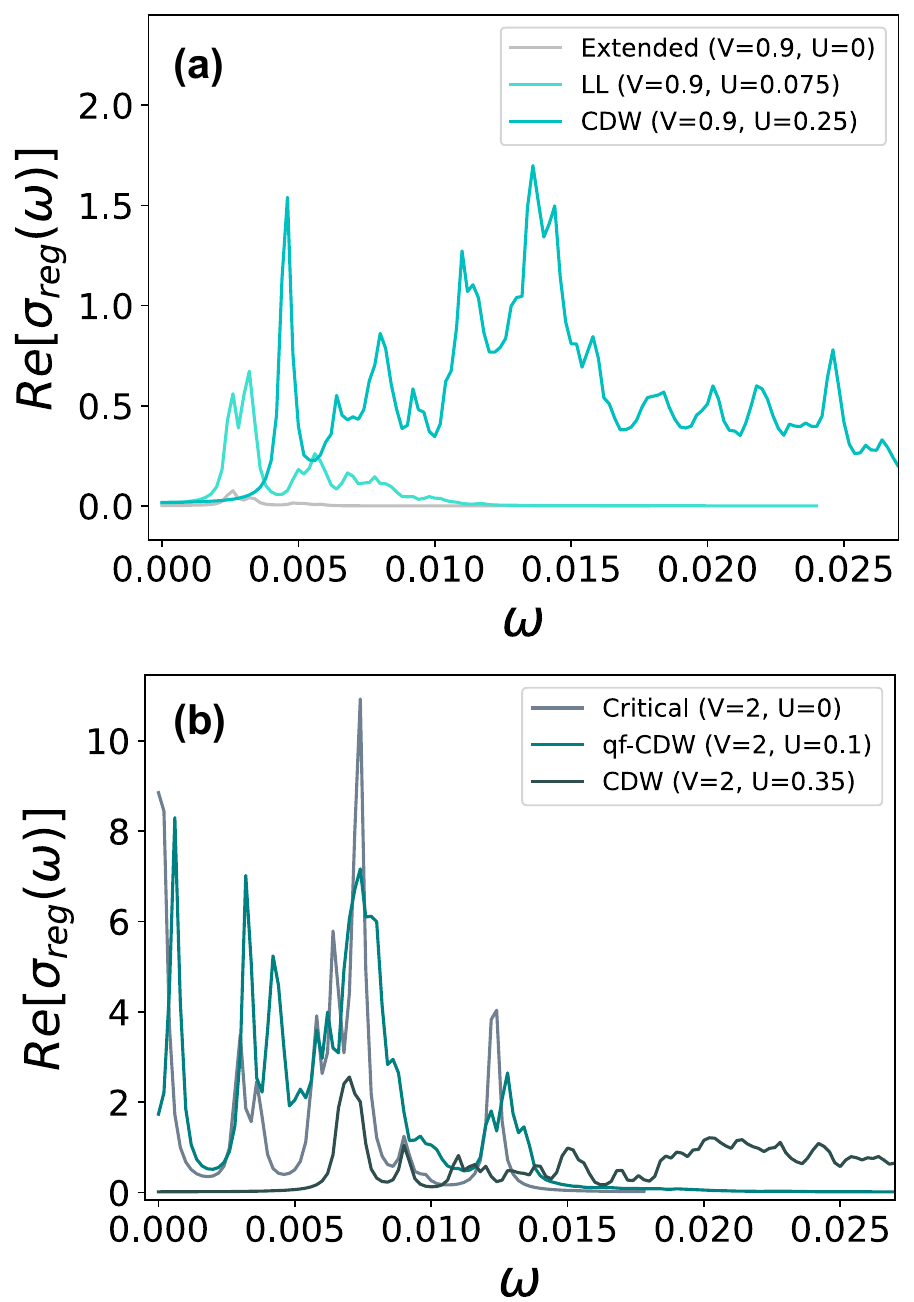}
\caption{Real component of the regular part of the optical conductivity, $\mathfrak{Re}[\sigma_{reg}(\omega)]$,
Eq. \ref{eq:11}, computed for $N=94$, with $\delta\omega=2\times10^{-4}$,
Lorentzian width of $\eta=$ $2\times10^{-4}$ and random shift and
twist, for: (a) extended states ($V_{2}=0.9,U=0$), LL ($V_{2}=0.9,U=0.075$)
and CDW ($V_{2}=0.9,U=0.25$); (b) critical states ($V_{2}=2,U=0$),
qf-CDW ($V_{2}=2$, $U=0.1$) and CDW ($V_{2}=2$, $U=0.35$). \protect\label{fig:5}}
\end{figure}

We now examine the dynamical response of the system through the regular
part of the optical conductivity, $\sigma_{\mathrm{reg}}(\omega)$,
focusing on frequencies well below the interband gap where intraband
processes dominate. Figure$\,$\ref{fig:5}(a) shows $\mathrm{Re},\sigma_{\mathrm{reg}}(\omega)$
for $V_{2}=0.9$.

In the non-interacting ($U=0$) extended phase, the regular part of
the conductivity is strongly suppressed at low frequencies. In this
regime, the quasiperiodic system is well approximated by a commensurate
superlattice with a large unit cell, and the narrow band near $E=0$
behaves effectively as a Bloch band. Consequently, intraband contributions
are entirely captured by the Drude peak at $\omega=0$, while $\sigma_{\mathrm{reg}}(\omega)$
contains only interband processes and vanishes at small $\omega$
(see Appendix \ref{sec:B}).

As $V_{2}$ increases toward and beyond unity, spectral gaps open
increasingly close to the Fermi energy, enabling low-energy intraband
transitions. For $V_{2}\ge1$, the Drude weight vanishes in the thermodynamic
limit and the spectral weight is transferred to the regular part of
the conductivity, resulting in a pronounced low-frequency response
even in the non-interacting critical phase {[}Fig.$\,$\ref{fig:5}(b){]}.

Turning on interactions at $V_{2}=0.9<1$ enhances the low-frequency
optical response in the LL phase, as expected from Luttinger-liquid
theory, where $\sigma(\omega)\sim\omega^{3}$ at small $\omega$ \cite{giamarchi1992conductivity}.
In contrast, CDW phase ($U=0.25$) phase exhibits a well-defined optical
gap \cite{feng2021signatures,shao2021using,fehske2006quantum} {[}see
Fig.$\,$\ref{fig:5}(a){]}, reflecting the finite energy required
to excite particle--hole pairs across the charge gap.

In the strongly quasiperiodic regime ($V_{2}\gtrsim2$), the non-interacting
critical phase already displays a strong optical response at low frequencies.
Upon introducing weak interactions ($U=0.1$), the system enters the
qf-CDW regime, which remains characterized by an almost vanishing
optical gap and a substantial low-frequency conductivity. Increasing
$U$ further drives the system into the CDW phase, where a clear optical
gap re-emerges, mirroring the behavior observed at $V_{2}=0.9$.

\begin{table}[h]
\begin{tabular}{|c|c|}
\hline 
Phase &
$K(\Omega)$\tabularnewline
\hline 
\hline 
Extended ($V=0.9,$$\ U=0$) &
$6\times10^{-5}$\tabularnewline
\hline 
LL ($V=0.9,$$\ U=0.075$) &
$9\times10^{-4}$\tabularnewline
\hline 
CDW ($V=0.9,$$\ U=0.25$) &
$1.1\times10^{-2}$\tabularnewline
\hline 
Critical ($V=2,$$\ U=0$) &
$1.6\times10^{-2}$\tabularnewline
\hline 
qf-CDW ($V=2,$$\ U=0.1$) &
$2.3\times10^{-2}$\tabularnewline
\hline 
CDW ($V=2,$$\ U=0.35$) &
$2.1\times10^{-2}$\tabularnewline
\hline 
\end{tabular}

\caption{Spectral weight of the regular part of the optical conductivity for
the intraband transitions in the flatband.\protect\label{tab:2}}
\end{table}

Additional insight is obtained by considering the integrated spectral
weight of the regular part of the conductivity, $K(\Omega)$, defined
in Eq.$\,$\ref{eq:13}. The values of $K(\Omega)$ for different
phases are summarized in Table \ref{tab:2}.

In the extended non-interacting phase, $K(\Omega)$ is extremely small,
reflecting the fact that nearly all spectral weight resides in the
Drude peak. By contrast, both CDW and qf-CDW phases exhibit significantly
larger values of $K(\Omega)$, of order $10^{-2}$, indicating that
the dominant contribution to the optical response arises from finite-frequency
processes. The LL phase occupies an intermediate regime, with a spectral
weight larger than in the extended phase but still substantially smaller
than in the ordered phases.

Overall, these results show that the redistribution of spectral weight
between the Drude peak and the regular part of the optical conductivity
provides a clear dynamical fingerprint of the different phases induced
by the interplay of quasiperiodicity and interactions.

\section{Conclusions}

In this work, we have demonstrated that band projection can provide
an accurate and efficient description of interacting quasiperiodic
systems, provided that virtual processes involving remote bands are
properly taken into account. Using a one-dimensional narrow-band quasiperiodic
model as a test bed, we showed that a projection restricted to first
order in the interaction strength fails to reproduce the correct ground-state
phase diagram. This breakdown contrasts with the widespread success
of first-order band projection in translationally invariant flatband
systems and highlights a fundamental difference introduced by quasiperiodicity.

By incorporating second-order virtual interband processes, we obtained
an effective Hamiltonian that quantitatively reproduces all phases
previously identified using numerically exact DMRG calculations. In
particular, the second-order band-projected theory captures the interaction-driven
transition between the Luttinger liquid and charge-density-wave phases
in the extended regime, as well as the crossover between the quasifractal
charge-density-wave regime and the conventional CDW phase in the strongly
quasiperiodic regime. Our results identify a specific class of second-order
processes as being qualitatively essential for stabilizing correlated
phases in narrow-band quasiperiodic systems.

Beyond ground-state properties, the controlled band projection scheme
allowed us to investigate low-energy neutral excitations and dynamical
response functions that are difficult to access with DMRG. We showed
that the Luttinger liquid phase is characterized by gapless neutral
and optical excitations, while the CDW phase exhibits finite neutral
and optical gaps. In contrast, the quasifractal CDW regime displays
strongly suppressed---though finite---neutral, charge, and optical
gaps, together with a redistribution of optical spectral weight toward
low frequencies. These features provide clear dynamical signatures
distinguishing the different interaction-driven regimes.

Our findings establish band projection, when extended beyond first
order, as a reliable tool for studying correlated quasiperiodic systems.
More broadly, they suggest a viable and controlled route to explore
interacting narrow-band quasiperiodic systems in higher dimensions,
where exact numerical methods are severely limited. This is particularly
relevant for moiré materials, such as twisted bilayer or trilayer
\cite{xia2025magic,uri2023superconductivity} graphene and twisted
bilayer graphene on hBN \cite{lai2025moire}, where quasiperiodicity
and narrow-bands naturally coexist and where the role of interactions
in the absence of strict translational symmetry remains an open and
timely problem.

\section*{Acknowledgments}

The authors MG and PR acknowledge partial support from Fundação para
a Ciência e Tecnologia (FCT-Portugal) through Grant No. UID/CTM/04540/2019. FR,
MG and PR acknowledge support by FCT through Grant No. UIDB/04540/2020. BA
and EVC acknowledge partial support from FCT- Portugal through Grant
No. UIDB/04650/2020. MG acknowledges further support from FCT-Portugal
through the Grant SFRH/BD/145152/2019. BA acknowledges further support
from FCT-Portugal through Grant No. CEECIND/02936/2017/CP1458/CT0010.
We finally acknowledge the Tianhe-2JK cluster at the Beijing Computational
Science Research Center (CSRC) and the OBLIVION supercomputer, through
projects 2022.15910.CPCA.A1 and 2023.14361.CPCA.A1 (based at the High
Performance Computing Center - University of Évora) funded by the
ENGAGE SKA Research Infrastructure (reference POCI-01-0145-FEDER-022217
- COMPETE 2020 and the Foundation for Science and Technology, Portugal)
and by the BigData@UE project (reference ALT20-03-0246-FEDER-000033
- FEDER and the Alentejo 2020 Regional Operational Program. Computer
assistance was provided by CSRC's and OBLIVION's support teams.

\pagebreak{}

\section*{Appendix A: Band projection technique in quasiperiodic systems \protect\label{sec:A}}

\renewcommand{\thefigure}{A\arabic{figure}}
\setcounter{figure}{0}

\renewcommand{\theequation}{A\arabic{equation}}
\setcounter{equation}{0}

The band projection technique is an effective framework for studying
many-body systems characterized by a flatband. This approach enables
computational analysis of significantly larger system sizes than would
be feasible using conventional exact diagonalization applied to the
full system.

Here we describe the detailed implementation of this technique, taking
the example considered in this work of a narrow-band quasi-periodic
system with the Hamiltonian given in Eq.$\,$\ref{eq:1}. At $V_{2}=1$,
a narrow-band (FB) is energetically separated from the remote bands.
For interactions that are weak when compared to the band gap ($\Delta$)
but strong when compared to the bandwidth ($W$), $W\ll U\ll\Delta,$
interaction-induced mixing between the FB and the lower (or upper)
bands is weak. In this regime, it is therefore well justified to account
for the mixing between the narrow-band subspace and the remote bands
perturbatively \cite{bergholtz2013topological,parameswaran2013fractional}.

Working in the band basis introduced in Eq.$\,$\ref{eq:2}, the single-particle
Hamiltonian is diagonal and given by $H_{0}=\sum_{i}\varepsilon_{i}d_{i}^{\dagger}d_{i}$,
where $\epsilon_{i}$ are the single-particle eigenenergies:

\begin{align}
H_{0} & =-\sum_{r}t_{r}\left(c_{r}^{\dagger}c_{r+1}+c_{r+1}^{\dagger}c_{r}\right)\nonumber \\
H_{0} & =-\sum_{i}\sum_{r}t_{r}\left(\langle i|r\rangle\langle r+1|i\rangle+\langle i|r+1\rangle\langle r|i\rangle\right)d_{i}^{\dagger}d_{i}\nonumber \\
H_{0} & =\sum_{i}\varepsilon_{i}d_{i}^{\dagger}d_{i}.
\end{align}

Since we work at half-filling of the narrow-band and consider weak
interactions when compared with the bandgap, we can assume that the
sates on the remote bands are ``frozen''. In particular, we assume
that the eigenstates in the lower-energy and higher-energy remote
bands are respectively occupied and empty, only renormalizing the
narrow-band through virtual processes.

The interacting part of the Hamiltonian in Eq.$\,$\ref{eq:1} can
be written as:
\begin{align}
V & =U\sum_{r}\left[\sum_{\substack{i,j,\\
m,n
}
}\langle i|r\rangle d_{i}^{\dagger}\langle j|r+1\rangle d_{j}^{\dagger}\langle r+1|m\rangle d_{m}\langle r|n\rangle d_{n}\right]\nonumber \\
 & =U\sum_{r}\left[\sum_{\substack{i,j,\\
m,n
}
}\langle i|r\rangle\langle j|r+1\rangle\langle r+1|m\rangle\langle r|n\rangle d_{i}^{\dagger}d_{j}^{\dagger}d_{m}d_{n}\right]\nonumber \\
 & =\sum_{\substack{i,j,\\
m,n
}
}V_{ijmn}^{A}d_{i}^{\dagger}d_{j}^{\dagger}d_{m}d_{n},
\end{align}
where $V_{ijmn}^{A}=U/4\left[V_{ijmn}-V_{jimn}-V_{ijnm}+V_{jinm}\right]$
are the antisymetrized interaction matrix elements $V_{ijmn}=\sum_{r}\langle i|r\rangle\langle j|r+1\rangle\langle r+1|m\rangle\langle r|n\rangle$.

By taking the OC and E eigenstates as fully occupied and empty, i.e.,
$d_{i}^{\dagger}d_{j}\rightarrow\langle d_{i}^{\dagger}d_{j}\rangle=\delta_{ij}\,\,,i,j\in\textrm{OC}$
and $d_{i}^{\dagger}d_{j}\rightarrow\langle d_{i}^{\dagger}d_{j}\rangle=0\,\,,i,j\in\textrm{E}$,
the projected Hamiltonian to first order in interaction strength is
given by

\begin{align}
H_{BP}^{(1)} & =H_{0}+H_{FB-OC}+H_{FB-FB}=\sum_{i\in hf}\varepsilon_{i}d_{i}^{\dagger}d_{i}+\nonumber \\
 & 4\sum_{\substack{i,l\ \in\ FB\\
j\ \in\ OC
}
}V_{ijjn}^{A}d_{i}^{\dagger}d_{j}+\sum_{\substack{i,j\ \in\ FB\\
m,n\ \in\ FB
}
}V_{ijmn}^{A}d_{i}^{\dagger}d_{j}^{\dagger}d_{m}d_{n},\label{eq:7-1}
\end{align}
where $H_{FB-OC}$ destroys and creates one particle in the OC band
and another particle in the FB, and $H_{FB-FB}$ is the projected
interaction in the FB subspace.

If the first-order approximation in Eq. \ref{eq:8-1} fails to capture
the expected behavior in the $U\ll\Delta$ regime, which is the case
in our model, second-order perturbation corrections must be included.
To proceed with the derivation, we adopt a more straightforward notation.
Operators will be written as $o^{\dagger},o;p^{\dagger},p;e^{\dagger},e;$
referring to the OC band (o); partially filled band/FB (p); and E
band (e). Operator indices will be denoted with a bottom bar, no bar
and top bar, to indicate they belong to the occupied band, partially
filled band and empty band, respectively. We also use Einstein's notation
in what follows. Eq. \ref{eq:7-1} can therefore be more compactly
written as:

\begin{align}
H_{BP}^{(1)} & =H_{0}+H_{FB-OC}+H_{FB-FB}\nonumber \\
 & =\varepsilon_{i}p_{i}^{\dagger}p_{i}+4V_{i\underline{o}\underline{o}n}^{A}p_{i}^{\dagger}p_{n}+V_{ijmn}^{A}p_{i}^{\dagger}p_{j}^{\dagger}p_{m}p_{n}.
\end{align}

To introduce BP up to second order, we first rewrite $H_{BP}^{(1)}$
as:

\begin{equation}
H_{BP}^{(1)}=P_{0}H_{0}P_{0}+P_{0}VP_{0}
\end{equation}
where $H_{0}$ and $V$ are defined in Eq. \ref{eq:1-1}and $P_{0}$
is the projection operator into the narrow-band eigenstates, $P_{0}=|\psi_{0}\rangle\langle\psi_{0}|,$
where:

\begin{align*}
|\psi_{0}\rangle & =|OC\rangle|FB\rangle=\prod_{\underline{\alpha}}^{N_{OC}}o_{\underline{\alpha}}^{\dagger}\prod_{\beta}^{N_{FB}/2}p_{\beta}^{\dagger}|0\rangle.
\end{align*}

Recall that the OC and E bands do not influence the dimensionality
of $H_{BP}$, since the states are either completely full (OC) or
completely empty (E): $|111\cdots111\rangle|FB\rangle|000\cdots000\rangle$.

Introducing $\overline{P}=\mathbb{I}-P_{0}$, the full Hamiltonian
can be written exactly as

\begin{equation}
\boldsymbol{H}=\left(\begin{array}{cc}
P_{0}HP_{0} & P_{0}H\overline{P}\\
\overline{P}HP_{0} & \overline{P}H\overline{P}
\end{array}\right)=\left(\begin{array}{cc}
P_{0}HP_{0} & P_{0}V\overline{P}\\
\overline{P}VP_{0} & \overline{P}H\overline{P}
\end{array}\right),
\end{equation}
where we used $P_{0}H_{0}\overline{P}=0$, since $H_{0}$ is diagonal.

Now we consider the Green function $\mathcal{G}=(E-\boldsymbol{H})^{-1}$
and define our effective projected Hamiltonian as $P_{0}\mathcal{G}P_{0}=(E-\boldsymbol{H}_{\boldsymbol{BP}})^{-1}.$
Explicitly, we have:

\begin{widetext}

\begin{align}
\mathcal{G} & =(E-\boldsymbol{H})^{-1}=\left(\begin{array}{cc}
P_{0}(E-H)P_{0} & -P_{0}V\overline{P}\\
-\overline{P}VP_{0} & \overline{P}(E-H)\overline{P}
\end{array}\right)^{-1}\nonumber \\
 & =\left(\begin{array}{cc}
\left(E-P_{0}HP_{0}-P_{0}V\overline{P}\frac{1}{\overline{P}(E-H)\overline{P}}\overline{P}VP_{0}\right)^{-1} & \cdots\\
\cdots & \cdots
\end{array}\right),\label{eq:14}
\end{align}

\end{widetext}which implies that we can write the effective Hamiltonian
up to second-order in interaction strength as:

\begin{equation}
H_{BP}^{(2)}=P_{0}H_{0}P_{0}+P_{0}VP_{0}+\frac{P_{0}V\overline{P}VP_{0}}{E-H_{0}}\,,
\end{equation}
where we used the fact that $\frac{1}{\overline{P}(E-H_{0})\overline{P}}$
is diagonal in the occupation number basis.

The non-zero contributions in first and second-order perturbation
theory can be represented using Feynman diagrams. These diagrams represent
the interactions between the flat, occupied, and empty bands, and
their combinatorial contributions are detailed in Fig.$\,$\ref{fig:2}.
For simplicity, we use a shorthand notation, e.g., $ee;pp$, to represent
$V_{\bar{i}\bar{j}kl}^{A}e_{\bar{i}}^{\dagger}e_{\bar{j}}^{\dagger}p_{k}p_{l}$
.

We now analyse the contributions to $P_{0}V\overline{P}VP_{0}$ separately.

\subsection*{Terms with flatband and empty band operators}

These terms can involve the excitation of one electron, Fig. \ref{fig:2}
(c) , or two electrons, Fig. \ref{fig:2} (d), from FB to E. For such
contributions, $P_{0}V\overline{P}VP_{0}$ can be written as:

\begin{equation}
\langle\psi_{0}|\left(\begin{array}{c}
2pp;ep\\
pp;ee
\end{array}\right)|\psi_{I}\rangle\frac{1}{E-H_{0}}\langle\psi_{I}|\left(\begin{array}{c}
2pe;pp\\
ee;pp
\end{array}\right)|\psi_{0}\rangle,\label{eq:16}
\end{equation}
where the factor 2 stems from considering all contributions, e.g.:

\begin{align}
p_{i}p_{j};e_{\bar{k}}p_{l}+p_{i}p_{j};p_{k}e_{\bar{l}} & =p_{i}p_{j};e_{\bar{k}}p_{l}-p_{i}p_{j};e_{\bar{l}}p_{k}\nonumber \\
 & =p_{i}p_{j};e_{\bar{k}}p_{l}+p_{i}p_{j};e_{\bar{k}}p_{l}\nonumber \\
 & =2p_{i}p_{j};e_{\bar{k}}p_{l}.
\end{align}

Note that terms that do not conserve particle number such as $(2pp;ep|eepp)$
are not allowed.

From Eq. \ref{eq:16} we obtain the following terms in second order:

\begin{align*}
4(pp;ep|pe;pp)\ \  & 3-body\\
(pp;ee|ee;pp)\ \  & 2-body
\end{align*}
To exemplify how such terms are explicitly evaluated, we take $pp;ep|pe;pp$,
which yields

\begin{widetext}

\begin{align}
4(pp;ep|pe;pp) & =4V_{ij\bar{k}l}^{A}V_{m\bar{n}rs}^{A}p_{i}^{\dagger}p_{j}^{\dagger}e_{\bar{k}}p_{l}\frac{1}{(-\Delta E_{rs}^{\bar{k}m})}p_{m}^{\dagger}e_{\bar{n}}^{\dagger}p_{r}p_{s}=4V_{ij\bar{k}l}^{A}V_{m\bar{n}rs}^{A}p_{i}^{\dagger}p_{j}^{\dagger}p_{l}\frac{1}{(-\Delta E_{rs}^{\bar{k}m})}p_{m}^{\dagger}p_{r}p_{s}e_{\bar{k}}e_{\bar{n}}^{\dagger}\nonumber \\
 & =4V_{ij\bar{k}l}^{A}V_{m\bar{n}rs}^{A}p_{i}^{\dagger}p_{j}^{\dagger}p_{l}\frac{1}{(-\Delta E_{rs}^{\bar{k}m})}p_{m}^{\dagger}p_{r}p_{s}\delta_{\overline{kn}}=\frac{4}{\Delta E_{rs}^{\bar{k}m}}V_{ij\bar{k}l}^{A}V_{m\bar{k}rs}^{A}p_{i}^{\dagger}p_{j}^{\dagger}p_{l}p_{m}^{\dagger}p_{r}p_{s}\nonumber \\
 & =\frac{4}{\Delta E_{0}^{\bar{k}}}V_{ij\bar{k}l}^{A}V_{m\bar{k}rs}^{A}\left[-p_{i}^{\dagger}p_{j}^{\dagger}p_{m}^{\dagger}p_{l}p_{r}p_{s}+p_{i}^{\dagger}p_{j}^{\dagger}p_{r}p_{s}\delta_{lm}\right]\nonumber \\
 & =\frac{4}{\Delta E_{0}^{\bar{s}}}V_{ij\bar{s}l}^{A}V_{k\bar{s}mn}^{A}\left[-p_{i}^{\dagger}p_{j}^{\dagger}p_{k}^{\dagger}p_{l}p_{m}p_{n}+p_{i}^{\dagger}p_{j}^{\dagger}p_{m}p_{n}\delta_{kl}\right]\nonumber \\
 & =\frac{4}{\Delta E_{0}^{\bar{s}}}\left[-V_{ij\bar{s}l}^{A}V_{k\bar{s}mn}^{A}p_{i}^{\dagger}p_{j}^{\dagger}p_{k}^{\dagger}p_{l}p_{m}p_{n}+V_{ij\bar{s}k}^{A}V_{k\bar{s}mn}^{A}p_{i}^{\dagger}p_{j}^{\dagger}p_{m}p_{n}\right],
\end{align}

\end{widetext}

where $\Delta E_{rs}^{\bar{k}m}$ corresponds to the single-particle
energy needed to destroy two particles at indices $r,s$ and create
them at indices $\overline{\ensuremath{k},}m$. Since, in our model,
all states on the FB have energy close to the Fermi Energy, we consider
$\Delta E_{0}^{\overline{s}}=-\epsilon_{\bar{s}}$, the difference
between $E_{F}=0$ and the state created at $\overline{\ensuremath{s}}$.

\subsection*{Terms with flatband and occupied band operators}

These terms correspond to the diagrams (e)-(i) in Fig.\ref{fig:2}.
For such contributions, $P_{0}V\overline{P}VP_{0}$ can be written
as:

\begin{equation}
\langle\psi_{0}|\left(\begin{array}{c}
2op;pp\\
oo;pp\\
2oo;po\\
2op;pp\\
2oo;po
\end{array}\right)|\psi_{I}\rangle\frac{1}{E-H_{0}}\langle\psi_{I}|\left(\begin{array}{c}
2pp;po\\
pp;oo\\
2pp;po\\
2op;oo\\
2op;oo
\end{array}\right)|\psi_{0}\rangle,\label{eq:18}
\end{equation}
where, once again, we ignore zero contributions, as well as contributions
that do not conserve particle number. From Eq. \ref{eq:18} we obtain
the following non-vanishing terms in second-order:

\begin{align*}
4(op;pp|pp;po)\ \  & 3-body\\
(oo;pp|pp;oo)\ \  & 2-body\\
4(oo;po|pp;po)\ \  & 2-body\\
4(op;pp|op;oo)\ \  & 2-body\\
4(oo;po|op;oo)\ \  & 1-body
\end{align*}

\subsection*{Terms with operators in all bands}

These terms are depicted by the diagrams (j)-(q) in Fig.\ref{fig:2}.
$P_{0}V\overline{P}VP_{0}$ can be written as:

\begin{equation}
\langle\psi_{0}|\left(\begin{array}{c}
4op;ep\\
4op;eo\\
2pp;ep\\
4op;eo\\
2oo;ep\\
2op;ee\\
2oo;eo\\
4op;ep
\end{array}\right)|\psi_{I}\rangle\frac{1}{E_{0}-\overline{H}_{0}}\langle\psi_{I}|\left(\begin{array}{c}
4pe;po\\
2pe;pp\\
4oe;po\\
4oe;po\\
2pe;oo\\
2ee;po\\
4pe;po\\
2oe;oo
\end{array}\right)|\psi_{0}\rangle,\label{eq:19}
\end{equation}

From Eq. \ref{eq:19} we obtain the following terms in second-order:

\begin{align*}
16(op;ep;pepo)\ \  & 2-body\\
8(op;eo|pe;pp)\ \  & 2-body\\
8(pp;ep|oe;po)\ \  & 2-body\\
16(op;eo|oe;po)\ \  & 1-body\\
4(oo;ep|pe;oo)\ \  & 1-body\\
4(op;ee|ee;po)\ \  & 1-body\\
8(oo;eo|pe;po)\ \  & 1-body\\
8(op;ep|oe;oo)\ \  & 1-body
\end{align*}

\subsection*{Full effective Hamiltonian}

The full second-order effective Hamiltonian is obtained by combining
all terms, after normal ordering:

\begin{widetext}

\begin{equation}
H_{BP}^{(2)}=\varepsilon_{i}p_{i}^{\dagger}p_{i}+V_{in}^{1b}p_{i}^{\dagger}p_{n}+V_{ijmn}^{2b}p_{i}^{\dagger}p_{j}^{\dagger}p_{m}p_{n}+V_{ijklmn}^{3b}p_{i}^{\dagger}p_{j}^{\dagger}p_{k}^{\dagger}p_{l}p_{m}p_{n},\label{eq:6-1}
\end{equation}
where we have explicitly:

\begin{align}
V_{in}^{1b} & =4V_{i\underline{rr}n}+\sum_{\text{second-order terms}}\nonumber \\
 & =4V_{i\underline{rr}n}+\frac{8}{\Delta E_{\underline{r}}^{0}}V_{\underline{r}ikl}V_{lkn\underline{r}}+\frac{8}{\Delta E_{\underline{rs}}^{00}}V_{\underline{rs}nl}V_{il\underline{sr}}+\frac{32}{\Delta E_{\underline{r}}^{0}}V_{\underline{sr}k\underline{s}}V_{kin\underline{r}}+\frac{16}{\Delta E_{\underline{r}}^{0}}V_{\underline{sr}n\underline{s}}V_{\underline{m}i\underline{m}\underline{r}}\nonumber \\
 & +\frac{16}{\Delta E_{0}^{\bar{s}}}V_{\underline{r}i\bar{s}l}V_{l\bar{s}n\underline{r}}-\frac{16}{\Delta E_{0}^{\bar{s}}}V_{\underline{r}i\bar{s}\underline{r}}V_{j\bar{s}mn}+\frac{16}{\Delta E_{0}^{\bar{s}}}V_{\underline{r}i\bar{s}\underline{r}}V_{\underline{m}\bar{s}n\underline{m}}+\frac{8}{\Delta E_{\underline{rs}}^{0\bar{k}}}V_{\underline{rs}\bar{k}n}V_{i\bar{k}\underline{rs}}\nonumber \\
 & +\frac{8}{\Delta E_{\underline{k}0}^{\overline{rs}}}V_{\underline{k}i\bar{r}\bar{s}}V_{\bar{sr}n\underline{k}}+\frac{32}{\Delta E_{r}^{\overline{k}}}V_{\underline{rs}\bar{k}\underline{s}}V_{i\bar{k}n\underline{r}}.
\end{align}

\begin{align}
V_{ijmn}^{2b} & =V_{ijmn}+\sum_{\text{second-order terms}}\nonumber \\
 & =V_{ijmn}+\frac{4}{\Delta E_{0}^{\bar{s}}}V_{ij\bar{s}k}V_{k\bar{s}mn}+\frac{2}{\Delta E_{00}^{\overline{rs}}}V_{ij\bar{r}\bar{s}}V_{\bar{s}\bar{r}mn}+\frac{16}{\Delta E_{\underline{r}}^{0}}V_{\underline{r}iml}V_{jln\underline{r}}+\frac{2}{\Delta E_{\underline{rs}}^{00}}V_{\underline{rs}mn}V_{ij\underline{sr}}\nonumber \\
 & +\frac{16}{\Delta E_{\underline{r}}^{0}}V_{\underline{sr}m\underline{s}}V_{ijn\underline{r}}-\frac{16}{\Delta E_{\underline{r}}^{\bar{s}}}V_{\underline{r}i\bar{s}m}V_{j\bar{s}n\underline{r}}-\frac{16}{\Delta E_{0}^{\bar{s}}}V_{\underline{r}i\bar{s}\underline{r}}V_{j\bar{s}mn}.
\end{align}

\begin{equation}
V_{ijklmn}^{3b}=\sum_{\text{second-order terms}}=-\frac{4}{\Delta E_{0}^{\bar{s}}}V_{ij\bar{s}l}V_{k\bar{s}mn}+\frac{4}{\Delta E_{\underline{r}}^{0}}V_{\underline{r}ilm}V_{jkn\underline{r}}.
\end{equation}

\end{widetext}

$\negthickspace$

\pagebreak{} $\negthickspace$

\pagebreak{}

\section*{Appendix B: Benchmark of the optical conductivity \protect\label{sec:B}}

\renewcommand{\thefigure}{B\arabic{figure}}
\setcounter{figure}{0}

\renewcommand{\theequation}{B\arabic{equation}}
\setcounter{equation}{0}

In the implementation of the band projected optical conductivity,
given by Eq. \ref{eq:12}, we restricted our analysis to operators
within the FB (see Eq. \ref{eq:11-1}). For many-body systems up to
second-order in perturbation theory, including operators from the
other bands, $e.g.$, destroying a state in the occupied (OC) band
and creating one in the FB, would demand performing exact diagonalization
(ED) for every state annihilated in the OC band, which is computationally
unfeasible for larger system sizes.

\begin{widetext}

\begin{figure}[H]
\centering{}\includegraphics[scale=0.33]{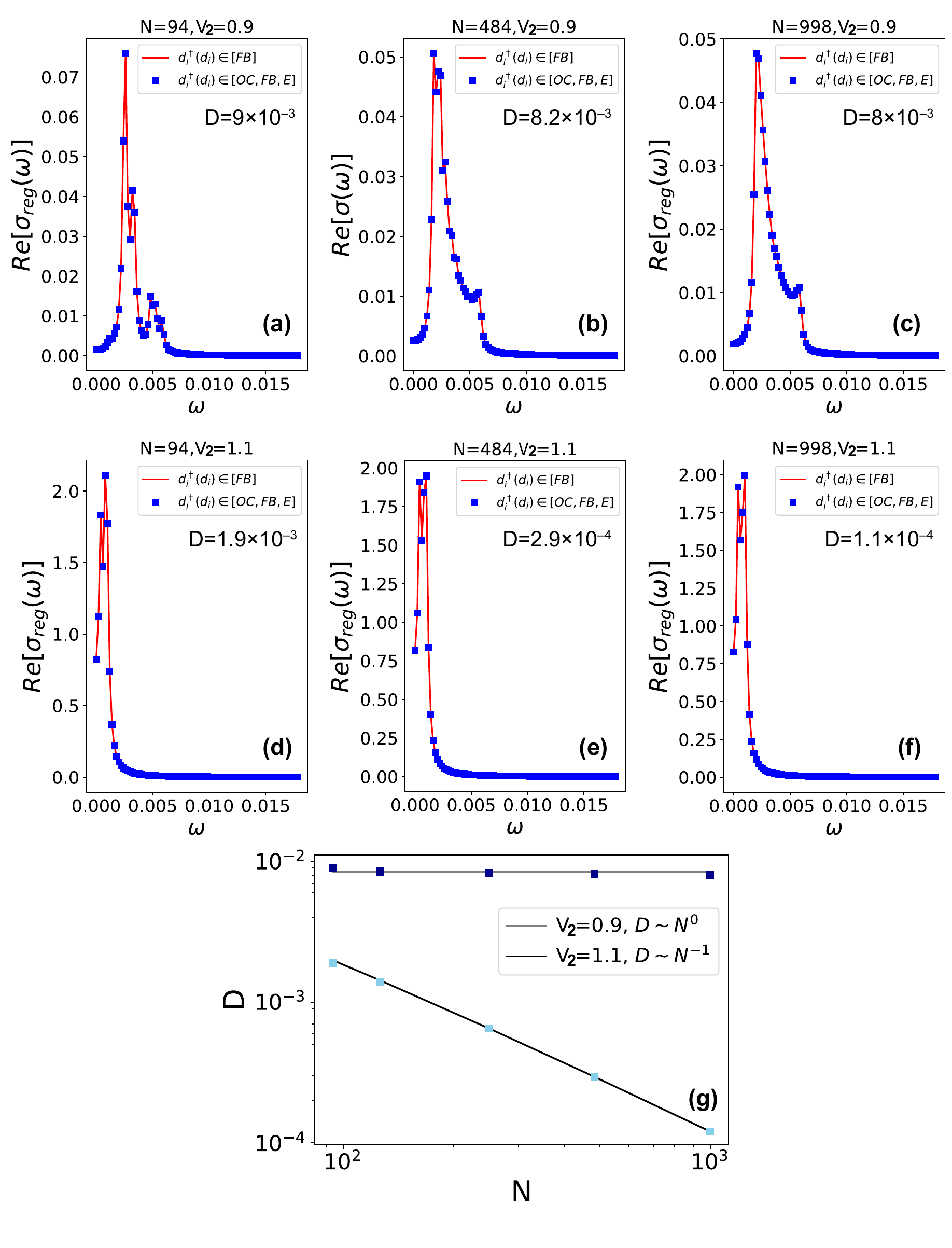}\caption{Real component of the finite-frequency optical conductivity in the
far infrared regime, computed with operators restricted to the FB
(red curves) and operators applied to all bands (blue squares). Interaction
strength is set to zero to make the computation with the full operators
feasible. Results for extended (critical) states are shown in $[a]-[c]$($[d]-[f]$),
with $\eta=$ $2\times10^{-4}$. Panel $[g]$ shows the scaling of
the Drude weight for extended ($D\sim N^{0}$) and critical ($D\sim N^{-1}$)
states. \protect\label{fig:C1}}
\end{figure}

\end{widetext}

Here we show that, in the far-infrared regime ($\omega\ll1$) at half-filling,
the optical conductivity can be accurately captured by limiting the
operators to the FB. Figure \ref{fig:C1} presents the benchmarking
of the optical conductivity in the single-particle picture, computed
with the operators restricted to the FB (red curves) and with the
operators acting also in the remote bands (blue squares). The results
are shown for extended ($V_{2}=0.9$, $[a]-[c]$) and critical states
($V_{2}=1.1$, $[d]-[f]$), for selected system sizes ($N=94,\ 484,\ 998$).
The value of the Drude weight is indicated in each of these panels.

As can be seen more clearly in panel $[g]$, while the Drude weight
has a finite value for extended states ($D\sim N^{0}$), it goes to
zero in the thermodynamic limit for critical states ($D\sim N^{-1}$).
Since for extended states the spectral weight is mainly concentrated
in the Drude peak, the regular component of the optical conductivity
displays reduced values, as expected.

\begin{figure}[bh]
\centering{}\includegraphics[scale=0.27]{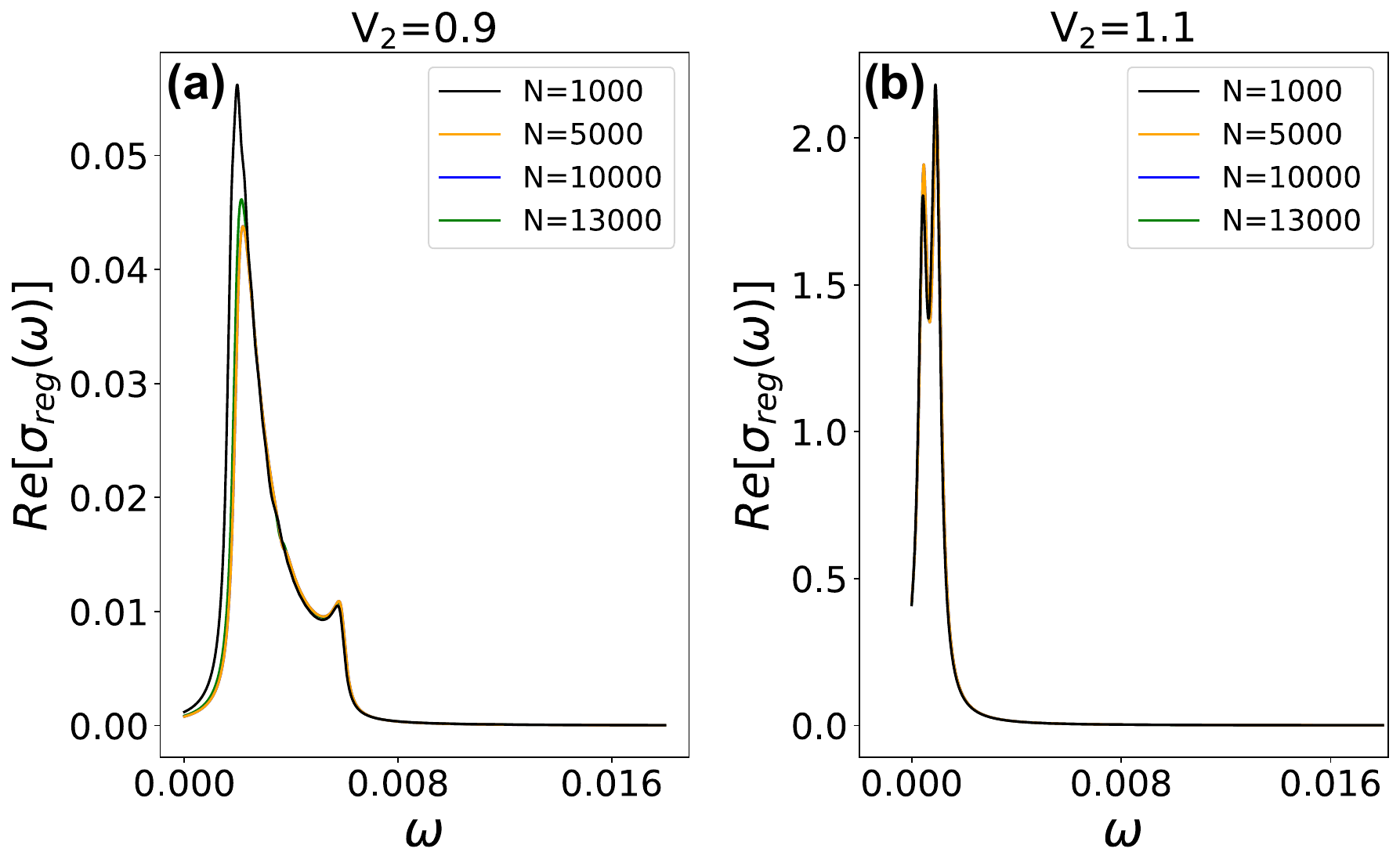}\caption{Real component of the finite-frequency optical conductivity in the
far infrared regime, for $N=[1000,\ 5000,\ 10000,\ 13000]$, with
$\eta=$ $2\times10^{-4}$. The results for extended states ($V_{2}=0..9$)
are shown in panel $[a]$ and the results for critical states ($V_{2}=1.1$)
in panel $[b]$. \protect\label{fig:C2}}
\end{figure}

Figure \ref{fig:C2} presents the optical conductivity for $N=[1000,\ 5000,\ 10000,\ 13000]$
and $V_{2}=[0.9,\ 1.1]$. The results for the different system sizes
are qualitative similar and converge as $N$ gets closer to the thermodynamic
limit, both for extended states and critical states .

\section*{Appendix C: Derivation and computation of the optical conductivity
with band projection \protect\label{sec:C}}

\renewcommand{\thefigure}{C\arabic{figure}}
\setcounter{figure}{0}

\renewcommand{\theequation}{C\arabic{equation}}
\setcounter{equation}{0}

The optical conductivity is a linear response function that relates
the current ($J$) to a transverse electric field ($E$)\cite{millis2004optical}:

\begin{equation}
J(\omega)=\sigma(\omega)E_{\perp}(\omega).\label{eq:1-1}
\end{equation}

The real part of the optical conductivity ($\sigma'$) is associated
with the energy absorption from the electric field, while the imaginary
part ($\sigma''$) represents the phase shift between the current
and the applied field, typical of phenomena such as polarization and
inductance. These terms are related by a Kramers-Kronig relation:

\begin{equation}
\sigma''(\omega)=\int_{-\infty}^{\ \infty}\frac{d\omega'}{\pi}\mathcal{P}\frac{\sigma'(\omega')}{\omega-\omega'},\label{eq:2-1}
\end{equation}

where $\mathcal{P}$ is the Cauchy principal value of the integral.
In general, one may write\cite{kirchner1999transport}:

\begin{equation}
\sigma(\omega)=\frac{iD}{\omega+i\eta}+\sigma_{reg}(\omega),\label{eq:3-1}
\end{equation}

where $D$ is the Drude weight and $\sigma_{reg}(\omega)$ is the
regular part of the optical conductivity, excluding the singular contribution
at $\omega=0.$

The Drude weight quantifies the response of free electrons to an applied
electric field\cite{takasan2023drude}. It can be used to characterize
ballistic transport and, more broadly, to identify Mott transitions.
For systems with discrete translational symmetry, $D$ can serve as
an order parameter: $D\neq0$ for metallic ground states and $D=0$
for non-metallic ground states\cite{millis2004optical}.

The Drude weight can be conveniently expressed in terms of a static
vector field ($A$)\cite{millis2004optical,kirchner1999transport}:

\begin{equation}
D=\frac{1}{N}\left.\frac{\partial^{2}E_{g.s.}}{\partial A^{2}}\right|_{A=0}\label{eq:4-1}
\end{equation}

For lattice model calculations, we write the Drude weight as:

\begin{equation}
D=\frac{1}{N}\left[\langle-T\rangle_{0}-2\sum_{\nu\neq0}\frac{|\langle\nu|J|0\rangle|^{2}}{E_{\nu}-E_{0}}\right],\label{eq:5-1}
\end{equation}
where $T$ and $J$ are the current and kinetic energy operators,
respectively given by:

\begin{align}
T & =-\sum_{r=0}^{N-1}t_{r}(c_{r}^{\dagger}c_{r+1}+c_{r+1}^{\dagger}c_{r}),\label{eq:8-1}\\
J & =\left.\frac{\partial H}{\partial A}\right|_{A=0}=-i\sum_{r=0}^{N-1}t_{r}(c_{r}^{\dagger}c_{r+1}-c_{r+1}^{\dagger}c_{r}),\label{eq:9-1}
\end{align}
and $t_{r}=1+Vcos[2\pi\tau(r+1/2)+\phi]$ is the hopping of our flatband
1D Hamiltonian, given by Eq.$\,$\ref{eq:1} in the main text.

Equation \ref{eq:5-1} can be obtained as follows: (i) couple the
model to an electromagnetic field via the Peierls substitution, $t_{r}\rightarrow t_{r}e^{i\boldsymbol{A}}$;
(ii) expand the Peierls phase up to second order, $e^{iA}\simeq1+i\boldsymbol{A-}\boldsymbol{A}^{2}/2$;
(iii) expand the energy shift between the ground states with and without
the vector field ($\boldsymbol{A}$), $\Delta E_{g.s.}=E_{0}(\boldsymbol{A})-E_{0}(0)$,
up to second order; (iv) compute $\Delta E_{g.s.}$ using second-order
perturbation theory.

To obtain the optical conductivity, we start with the Kubo formula:

\begin{equation}
\langle J_{total}\rangle(t)=\frac{1}{N}\left[\langle T\rangle_{0}A(t)+\int dt'\Pi(t,t')A(t')\right],\label{eq:14-1}
\end{equation}
where the current-current correlation function is:

\begin{equation}
\Pi(t,t')=-i\Theta(t-t')\langle[J(t),J(t')]\rangle,\label{eq:15}
\end{equation}
and the step function, $\Theta(t-t')$, ensures causality.

In the Lehmann representation, we write:

{\scriptsize
\begin{align}
\Pi(t,t') & =-i\Theta(t-t')\frac{1}{Z}\sum_{\alpha,\beta}\left(e^{-\beta E_{\alpha}}-e^{-\beta E_{\beta}}\right)e^{i(E_{\alpha}-E_{\beta})(t-t')}J_{\alpha\beta}J_{\beta\alpha}\nonumber \\
\Pi(t,t') & =-i\Theta(t-t')\frac{1}{Z}\sum_{\alpha\neq\beta}\left(e^{-\beta E_{\alpha}}-e^{-\beta E_{\beta}}\right)e^{i(E_{\alpha}-E_{\beta})(t-t')}J_{\alpha\beta}J_{\beta\alpha},\label{eq:16-1}
\end{align}
}where $J_{\alpha\beta}=\langle\alpha|J|\beta\rangle$. Notice that
when $\alpha=\beta$, $\Pi(t,t')=0$ due to the term $\left(e^{-\beta E_{\alpha}}-e^{-\beta E_{\beta}}\right)$,
since $J_{\alpha\alpha}$ is finite.

To move to the frequency domain, we perform a Fourier transform:

\begin{align}
\Pi(\omega) & =\int_{\ 0}^{\ \infty}dt\ e^{i\omega t}\Pi(t,0).\label{eq:17}
\end{align}

Using the integral representation of the step function:

\begin{equation}
\Theta(t)=-\int_{\ -\infty}^{\ \ \infty}\frac{d\omega}{2\pi i}\frac{e^{-i\omega t}}{\omega+i\eta},\label{eq:18-1}
\end{equation}
we find:

\begin{widetext}

\begin{align}
\int_{\ 0}^{\ \infty}dt\ e^{i\omega t}\Theta(t)e^{i(E_{\alpha}-E_{\beta})t} & =-\int_{\ 0}^{\ \infty}dt\ e^{i\omega t}e^{i(E_{\alpha}-E_{\beta})t}\int_{\ -\infty}^{\ \ \infty}\frac{d\omega^{\prime}}{2\pi i}\frac{e^{-i\omega^{\prime}t}}{\omega^{\prime}+i\eta}\nonumber \\
 & =i\int_{\ -\infty}^{\ \ \infty}\frac{d\omega^{\prime}}{\omega^{\prime}+i\eta}\int_{\ 0}^{\ \infty}\frac{dt}{2\pi}e^{i(\omega+E_{\alpha}-E_{\beta}-\omega^{\prime})t}\nonumber \\
 & =i\int_{\ -\infty}^{\ \ \infty}\frac{d\omega^{\prime}}{\omega^{\prime}+i\eta}\delta(\omega+E_{\alpha}-E_{\beta}-\omega^{\prime})=i\frac{1}{\omega+E_{\alpha}-E_{\beta}+i\eta}.\label{eq:19-1}
\end{align}
\end{widetext}

This representation of the step function naturally introduces the
infinitesimal term $i\eta$ avoiding the need for \textit{ad hoc}
adjustments to ensure convergence of the integral.

Plugging this result into Eq.\ref{eq:17}, we have

\begin{equation}
\Pi(\omega)=\frac{1}{Z}\sum_{\alpha\neq\beta}\left(e^{-\beta E_{\alpha}}-e^{-\beta E_{\beta}}\right)\frac{J_{\alpha\beta}J_{\beta\alpha}}{\omega+E_{\alpha}-E_{\beta}+i\eta}.\label{eq:20}
\end{equation}
At zero temperature, the nonzero contributions arise when $|\alpha\rangle$
corresponds to the ground state and $|\beta\rangle$ to the excited
states or vice-versa. Therefore, we have

\begin{align}
\begin{cases}
\frac{\left(e^{-\beta E_{\alpha}}-e^{-\beta E_{\beta}}\right)}{Z} & =1\ \ \ \textrm{if}\ |\alpha\rangle=|0\rangle,\\
\frac{\left(e^{-\beta E_{\alpha}}-e^{-\beta E_{\beta}}\right)}{Z} & =-1\ \ \ \textrm{if}\ |\beta\rangle=|0\rangle,
\end{cases}\label{eq:21}
\end{align}

Substituting this into Eq. \ref{eq:20} gives us the current-current
correlation function at $T=0$:

\begin{align}
\Pi(\omega) & =\sum_{\nu\neq0}\frac{J_{0\nu}J_{\nu0}}{\omega+E_{0}-E_{\nu}+i\eta}-\frac{J_{0\nu}J_{\nu0}}{\omega-E_{0}+E_{\nu}+i\eta}\nonumber \\
 & =\sum_{\nu\neq0}\frac{J_{0\nu}J_{\nu0}}{\omega-\Delta E_{0}^{\nu}+i\eta}-\frac{J_{0\nu}J_{\nu0}}{\omega+\Delta E_{0}^{\nu}+i\eta},\label{eq:22}
\end{align}
where $\Delta E_{0}^{\nu}=E_{\nu}-E_{0}$.

Therefore, Eq. \ref{eq:14-1} becomes:

\begin{equation}
\langle J_{total}\rangle(\omega)=\frac{1}{N}\left[\langle T\rangle_{0}+\Pi(\omega)\right]\boldsymbol{A}(\omega).\label{eq:23}
\end{equation}

The conductivity in the frequency domain is defined as:

\begin{equation}
\langle J_{total}\rangle(\omega)=\sigma(\omega)\boldsymbol{E}(\omega),\label{eq:24}
\end{equation}
where $\boldsymbol{E}(\omega)=-$$i(\omega+i\eta)\boldsymbol{A}(\omega)$.
Substituting Eq. \ref{eq:24} into Eq. \ref{eq:19-1}:

\begin{align}
\sigma(\omega) & i(\omega+i\eta)=\frac{1}{N}\left[\langle T\rangle_{0}+\Pi(\omega)\right],\nonumber \\
\leftrightarrow\sigma(\omega) & =\frac{i}{\omega+i\eta}\left[\frac{\langle-T\rangle_{0}}{N}+\frac{1}{N}\sum_{\nu\neq0}\frac{|\langle\nu|J|0\rangle|^{2}}{\omega-\Delta E_{0}^{\nu}+i\eta}\right.\nonumber \\
 & \left.-\frac{1}{N}\sum_{\nu\neq0}\frac{|\langle\nu|J|0\rangle|^{2}}{\omega+\Delta E_{0}^{\nu}+i\eta}\right].\label{eq:25}
\end{align}

To express the optical conductivity in terms of the Drude weight and
its regular part, we combine Eqs. \ref{eq:5-1} and \ref{eq:25} to
get

{\small
\begin{align}
\sigma(\omega) & =\frac{i}{\omega+i\eta}\left[D+\frac{2}{N}\sum_{\nu\neq0}\frac{|\langle\nu|J|0\rangle|^{2}}{\Delta E_{0}^{\nu}}\right.\nonumber \\
 & \left.+\frac{1}{N}\sum_{\nu\neq0}\frac{|\langle\nu|J|0\rangle|^{2}}{\omega-\Delta E_{0}^{\nu}+i\eta}-\frac{1}{N}\sum_{\nu\neq0}\frac{|\langle\nu|J|0\rangle|^{2}}{\omega+\Delta E_{0}^{\nu}+i\eta}\right],\nonumber \\
\sigma(\omega) & =\frac{iD}{\omega+i\eta}+\sigma_{reg}(\omega),\label{eq:26}
\end{align}
}thereby recovering Eq. \ref{eq:3-1}, where the regular part of the
optical conductivity is given by:

{\small
\begin{align}
\sigma_{reg}(\omega) & =\frac{i}{\omega+i\eta}\left[\frac{2}{N}\sum_{\nu\neq0}\frac{|\langle\nu|J|0\rangle|^{2}}{\Delta E_{0}^{\nu}}\right.\nonumber \\
 & \left.+\frac{1}{N}\sum_{\nu\neq0}\frac{|\langle\nu|J|0\rangle|^{2}}{\omega-\Delta E_{0}^{\nu}+i\eta}-\frac{1}{N}\sum_{\nu\neq0}\frac{|\langle\nu|J|0\rangle|^{2}}{\omega+\Delta E_{0}^{\nu}+i\eta}\right].\label{eq:27}
\end{align}
}{\small\par}

It is important to note that $lim_{\omega\rightarrow0}\sigma_{reg}(\omega)\simeq\pi\delta(\omega)\times0=0$,
if we take $\eta{}^{2}\simeq0$. This result is expected for systems
with discrete translation invariance at zero temperature, but may
not hold when translational symmetry is broken \cite{millis1990interaction}.

To extract the real part of the optical conductivity, we first manipulate
the complex prefactor on the right-hand side of Eq. \ref{eq:27} to
separate its real and imaginary parts:

\begin{equation}
\frac{i}{\omega+i\eta}=\frac{i(\omega-i\eta)}{(\omega+i\eta)(\omega-i\eta)}=\frac{i\omega+\eta}{\omega^{2}+\eta^{2}}.\label{eq:28}
\end{equation}

We apply the same procedure to the middle term in the right-hand side
of Eq.\ref{eq:27}:

\begin{equation}
\frac{|\langle\nu|J|0\rangle|^{2}}{\omega-\Delta E_{0}^{\nu}+i\eta}=\frac{|\langle\nu|J|0\rangle|^{2}(\omega-\Delta E_{0}^{\nu}-i\eta)}{(\omega-\Delta E_{0}^{\nu})^{2}+\eta{}^{2}}.\label{eq:29}
\end{equation}

Analogously, for the last term:

\begin{equation}
\frac{|\langle\nu|J|0\rangle|^{2}}{\omega+\Delta E_{0}^{\nu}+i\eta}=\frac{|\langle\nu|J|0\rangle|^{2}(\omega+\Delta E_{0}^{\nu}-i\eta)}{(\omega+\Delta E_{0}^{\nu})^{2}+\eta^{2}}.\label{eq:30}
\end{equation}

Substituting Eqs. \ref{eq:28}, \ref{eq:29}, \ref{eq:30} into Eq.
\ref{eq:27}, we get:

\begin{align}
\sigma_{reg}(\omega) & =\frac{i\omega+\eta}{\omega^{2}+\eta^{2}}\left[\frac{2}{N}\sum_{\nu\neq0}\frac{|\langle\nu|J|0\rangle|^{2}}{\Delta E_{0}^{\nu}}\right.\nonumber \\
 & +\frac{1}{N}\sum_{\nu\neq0}\frac{|\langle\nu|J|0\rangle|^{2}(\omega-\Delta E_{0}^{\nu}-i\eta)}{(\omega-\Delta E_{0}^{\nu})^{2}+\eta^{2}}\\
 & \left.-\frac{1}{N}\sum_{\nu\neq0}\frac{|\langle\nu|J|0\rangle|^{2}(\omega+\Delta E_{0}^{\nu}-i\eta)}{(\omega+\Delta E_{0}^{\nu})^{2}+\eta^{2}}\right].\label{eq:31}
\end{align}

The real part of Eq.\ref{eq:31} is straightforward to derive and
is given by:

\begin{widetext}

\begin{align}
\mathfrak{Re}\left[\sigma_{reg}(\omega)\right] & =\frac{\eta}{\omega^{2}+\eta^{2}}\frac{2}{N}\sum_{\nu\neq0}\frac{|\langle\nu|J|0\rangle|^{2}}{\Delta E_{0}^{\nu}}+\frac{1}{\omega^{2}+\eta^{2}}\frac{1}{N}\sum_{\nu\neq0}|\langle\nu|J|0\rangle|^{2}\left[(\omega-\Delta E_{0}^{\nu})L_{(\omega,\Delta E_{0}^{\nu},\eta)}^{+}\right.\nonumber \\
 & \left.-(\omega+\Delta E_{0}^{\nu})L_{(\omega,\Delta E_{0}^{\nu},\eta)}^{-}\right]+\frac{\omega}{\omega^{2}+\eta^{2}}\frac{1}{N}\sum|\langle\nu|J|0\rangle|_{\nu\neq0}^{2}\left[L_{(\omega,\Delta E_{0}^{\nu},\eta)}^{+}-L_{(\omega,\Delta E_{0}^{\nu},\eta)}^{-}\right],\label{eq:32}
\end{align}

\end{widetext}

where the Lorentzian terms are:

\begin{align}
L_{(\omega,\Delta E_{0}^{\nu},\eta)}^{+} & =\frac{\eta}{(\omega-\Delta E_{0}^{\nu})^{2}+\eta^{2}},\label{eq:33}\\
L_{(\omega,\Delta E_{0}^{\nu},\eta)}^{-} & =\frac{\eta}{(\omega+\Delta E_{0}^{\nu})^{2}+\eta^{2}}.\label{eq:34}
\end{align}

Explicitly taking the limit $\eta\rightarrow0$, we have:

\begin{align}
\mathfrak{Re}\left[\sigma_{reg}(\omega)\right] & =\frac{\pi}{N}\sum_{\nu\neq0}\frac{|\langle\nu|J|0\rangle|^{2}}{\omega^{2}}\left[(\omega-\Delta E_{0}^{\nu})\delta_{(\omega-\Delta E_{0}^{\nu})}\right.\nonumber \\
 & \left.-(\omega+\Delta E_{0}^{\nu})\delta_{(\omega+\Delta E_{0}^{\nu})}\right]\nonumber \\
 & +\frac{\pi}{N}\sum_{\nu\neq0}\frac{|\langle\nu|J|0\rangle|^{2}}{\omega}\left[\delta_{(\omega-\Delta E_{0}^{\nu})}-\delta_{(\omega+\Delta E_{0}^{\nu})}\right],\nonumber \\
 & =\frac{\pi}{N}\sum_{\nu\neq0}\frac{|\langle\nu|J|0\rangle|^{2}}{\omega}\left[\delta_{(\omega-\Delta E_{0}^{\nu})}-\delta_{(\omega+\Delta E_{0}^{\nu})}\right],\nonumber \\
 & =\frac{\pi}{N}\sum_{\nu\neq0}\frac{|\langle\nu|J|0\rangle|^{2}}{\Delta E_{0}^{\nu}}\delta_{(|\omega|-\Delta E_{0}^{\nu})}.\label{eq:35}
\end{align}

The first two terms in the first line of Eq. \ref{eq:35} are null
since $x\delta(x)=0$.

Eq. \ref{eq:35} exhibits a symmetric structure with respect to $\omega.$
Given that $E_{0}^{\nu}>0$, the first delta function in the second-to-last
line of Eq. \ref{eq:35} is localized only for positive values of
$\omega$, while the second delta function is localized for negative
values. In the last line, the equation is presented in a form commonly
found in the literature\cite{fye1991drude,shastry1990twisted}.

The term $|\langle\nu|J|0\rangle|^{2}$ in Eq.$\,$\ref{eq:35} can
be written in terms of the FB operators as

\begin{widetext}

\begin{align}
|\langle\nu|J|0\rangle|^{2} & =\langle0|\sum_{j,k\in FB}-i\sum_{r=0}^{N-1}t_{r}\left[\langle j|r\rangle\langle r+1|k\rangle-\langle j|r+1\rangle\langle r|k\rangle\right]d_{j}^{\dagger}d_{k}|\nu\rangle\nonumber \\
 & \times\langle\nu|i\sum_{l,m\in FB}t_{r}\left[\langle l|r+1\rangle\langle r|m\rangle-\langle l|r\rangle\langle r+1|m\rangle\right]d_{l}^{\dagger}d_{m}|0\rangle.\label{eq:11-1}
\end{align}

\end{widetext}

\section*{Appendix D: First-order perturbation theory results \protect\label{sec:D}}

\renewcommand{\thefigure}{D\arabic{figure}}
\setcounter{figure}{0}

\renewcommand{\theequation}{D\arabic{equation}}
\setcounter{equation}{0}

\begin{figure}[h]
\centering{}\includegraphics[scale=0.23]{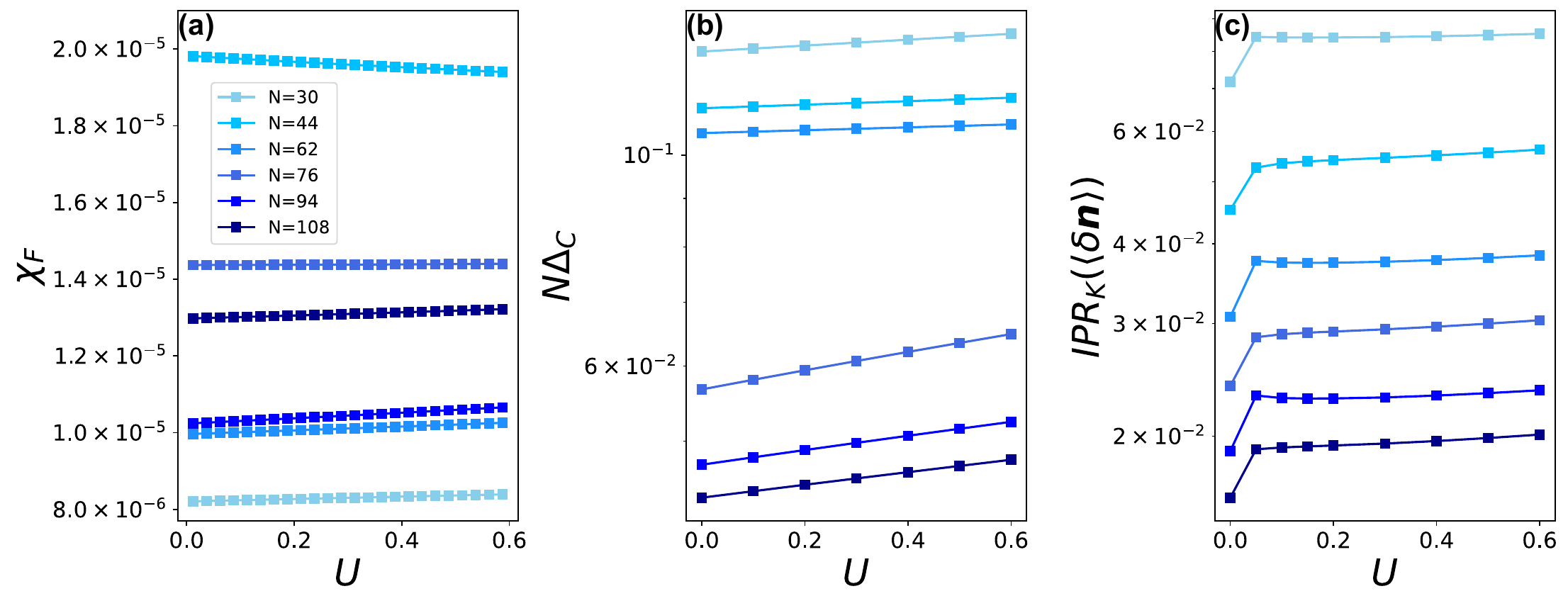}\caption{First order band projection for the computations of: (a) the fidelity
susceptibility ($\chi_{F}$), defined by Eq. \ref{eq:7} with $\delta U=1.25\times10^{-2}$;
(b) the charge gap ($\Delta_{C}$), defined by Eq. \ref{eq:8}; (c)
the IPR of the average of the density fluctuations in momentum space
($IPR_{K}(\langle\delta\boldsymbol{n}\rangle)$), defined by Eq. \ref{eq:9}.
Results are averaged over 10 random configurations of $\phi$ and
$\theta$.\protect\label{fig:D1}}
\end{figure}

Below we show the results for the fidelity susceptibility, the charge
gap and the IPR of the average of the density fluctuations in momentum
space using first order band projection, given by Eq. \ref{eq:4},
for $V_{2}=0.9$. As we can see in Fig. \ref{fig:D1}(a)-(c), no transition
is captured. The results for $\Delta_{C}$ and $IPR_{K}(\langle\delta\boldsymbol{n}\rangle)$
indicate that, even when increasing interactions, we still remain
in the LL phase. The results contrast drastically with those of Fig.
\ref{fig:3}(a)-(c), where we used band projection up to second order.

\begin{figure}[h]
\centering{}\includegraphics[scale=0.23]{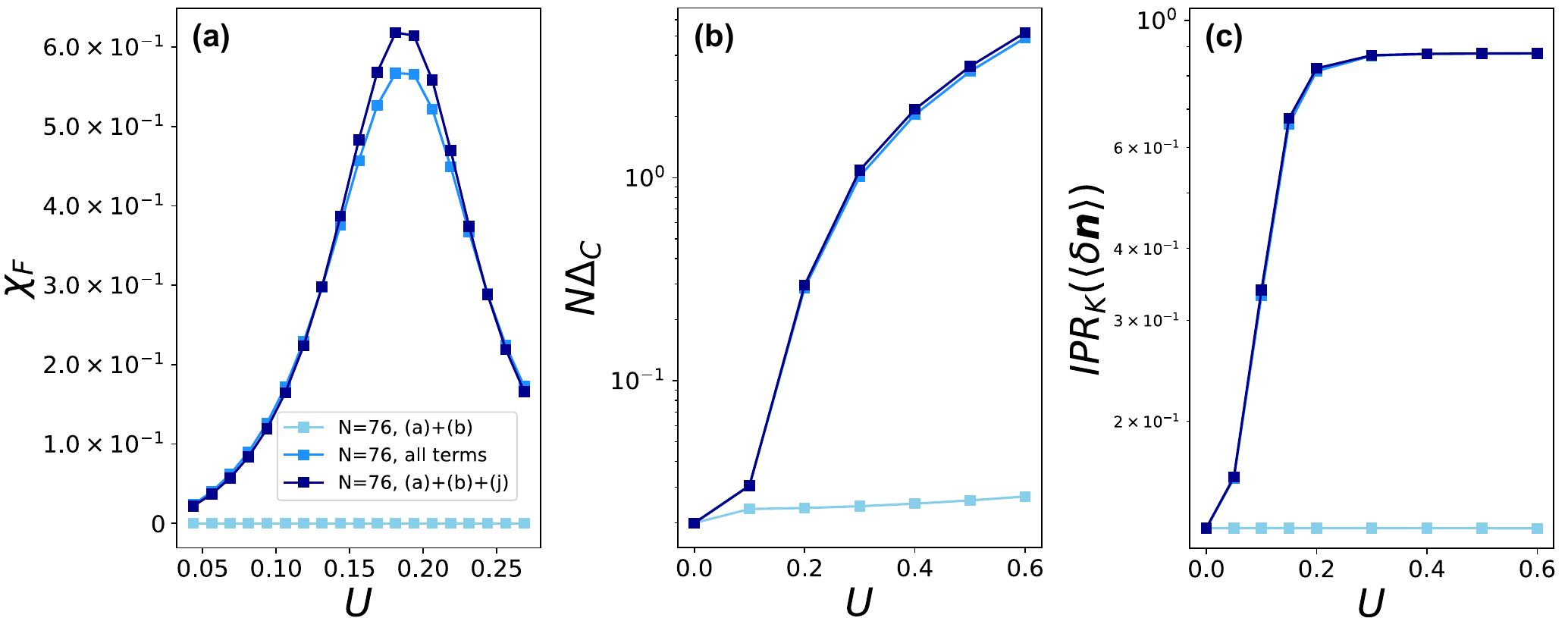}\caption{Computation of (a) the fidelity susceptibility ($\chi_{F}$), defined
by Eq. \ref{eq:7} with $\delta U=1.25\times10^{-2}$; (b) the charge
gap ($\Delta_{C}$), defined by Eq. \ref{eq:8}; (c) the IPR of the
average of the density fluctuations in momentum space ($IPR_{K}(\langle\delta\boldsymbol{n}\rangle)$),
defined by Eq. \ref{eq:9}, using first-order BP, BP up to second-order
and first-order terms plus the most relevant second-order term, for
$N=76$, $V_{2}=0.9$. \protect\label{fig:D2}}
\end{figure}

We also compare the results for $\chi_{F},$$\Delta_{C}$ and $IPR_{K}(\langle\delta\boldsymbol{n}\rangle)$,
Fig. \ref{fig:D2}(a)-(c), using first-order BP (light blue curves),
BP up to second order (blue curves) and first-order terms with the
most relevant second-order term (dark blue curves), corresponding
to the Feyman Diagrams (a), (b) and (j) in Fig. \ref{fig:2}. The
second-order contribution is a two-body term in the FB subspace, involving
contributions from all bands.

As can be seen in Fig. \ref{fig:D2}, the results for (a)+(b)+(j)
capture the relevant Physics of the model and are very close to those
from BP up to second order.

\pagebreak{}

\bibliographystyle{unsrtnat}
\bibliography{9C__Users_flavi_OneDrive_PhD_PHYSICS_artigos_band_projection_band_projection}

\end{document}